\documentclass{article}
\usepackage{epsf}
\usepackage{amssymb}

\newcommand{\epubtkAuthorData}[4]{#1\\#2\\email: #3\\\url{#4}}
\newcommand{\epubtkKeywords}[1]{}

\usepackage[dvips, breaklinks]{hyperref}
\usepackage{fancyvrb}

\begin{document}

\title{Characteristic Evolution and Matching}

\author{\epubtkAuthorData{Jeffrey Winicour}
        {Max-Planck-Institut f\" ur Gravitationsphysik \\
        Albert-Einstein-Institut \\
        Am M\"uhlenberg 1 \\
        14476 Golm, Germany \\
        and \\
        Department of Physics and Astronomy \\
        University of Pittsburgh \\
        Pittsburgh, PA 15260, U.S.A.}
       {jeff@einstein.phyast.pitt.edu}
       {http://www.phyast.pitt.edu/People/Faculty/J_Winicour.htm}}

\date{}
\maketitle


\begin{abstract}
  I review the development of numerical evolution codes for general
  relativity based upon the characteristic initial value
  problem. Progress is traced from the early stage of 1D feasibility
  studies to 2D axisymmetric codes that accurately simulate the
  oscillations and gravitational collapse of relativistic stars and to
  current 3D codes that provide pieces of a binary black hole
  spacetime. A prime application of characteristic evolution is to
  compute waveforms via Cauchy-characteristic matching, which is also
  reviewed.
\end{abstract}

\newpage


\section{Introduction}
\label{intro}

It is my pleasure to review progress in numerical relativity based
upon characteristic evolution. In the spirit of {\it Living Reviews in
  Relativity}, I invite my colleagues to continue to send me
contributions and comments at jeff@einstein.phyast.pitt.edu.

We are now in an era in which Einstein's equations can effectively be
considered solved at the local level. Several groups, as reported here and in
other {\it Living Reviews in Relativity}, have developed 3D codes which are stable and accurate in
some sufficiently local setting. Global solutions are another matter. In
particular, there is no single code in existence today which purports to be
capable of computing the waveform of gravitational radiation emanating from the
inspiral and merger of two black holes, the premier problem in classical
relativity. Just as several coordinate patches are necessary to describe a
spacetime with nontrivial topology, the most effective attack on the binary
black hole problem may involve patching together pieces of spacetime
handled by a combination of different codes and techniques.

Most of  the effort in numerical relativity has centered about the
Cauchy \{3\,+\,1\} formalism~\cite{york}, with the gravitational
radiation extracted by
perturbative methods based upon introducing an artificial Schwarzschild
background in the exterior
region~\cite{ab1,ab2,ab3,all1,rupright,rezzmatz,nagar}. These wave extraction
methods have not been thoroughly tested in a nonlinear 3D setting. A different
approach which is specifically tailored to study radiation is based upon the
characteristic initial value problem. In the 1960's, Bondi~\cite{1bondi,bondi}
and Penrose~\cite{Penrose}  pioneered the use of null hypersurfaces to describe
gravitational waves. This new approach has flourished in general relativity. It
led to the first unambiguous description of gravitational radiation in a fully
nonlinear context. It yields the standard linearized description of the
``plus'' and ``cross'' polarization modes of gravitational radiation in terms
of the Bondi news function $N$ at future null infinity $\mathcal{I}^+$. The Bondi
news function is an invariantly defined complex radiation amplitude $N=
N_{\oplus}+i N_{\otimes}$, whose real and imaginary parts correspond to the
time derivatives $\partial_t h_{\oplus}$ and $\partial_t h_{\otimes}$ of the
``plus'' and ``cross'' polarization modes of the strain $h$ incident on a
gravitational wave antenna.

The major drawback of the characteristic
approach arises from the formation of caustics in the light rays generating the
null hypersurfaces. In the most ambitious scheme proposed at the theoretical
level such caustics would be treated ``head-on'' as part of the evolution
problem~\cite{friedst1}. This is a profoundly attractive idea. Only a few
structural stable caustics can arise in numerical evolution, and their
geometrical properties are well enough understood to model their singular
behavior numerically~\cite{friedst2}, although a computational implementation
has not yet been attempted.

In the typical setting for the characteristic initial value problem, the domain
of dependence of a single nonsingular null hypersurface is empty. In order to
obtain a nontrivial evolution problem, the null hypersurface must either be
completed to a caustic-crossover region where it pinches off, or an additional
boundary must be introduced. So far, the only caustics that have been
successfully evolved numerically in general relativity are pure point caustics
(the complete null cone problem). When spherical symmetry is not present, it
turns out that the stability conditions near the vertex of a light cone place a
strong restriction on the allowed time step~\cite{igw}. Point caustics in
general relativity have been successfully handled this way for axisymmetric
spacetimes~\cite{papa}, but the computational demands for 3D evolution would
be prohibitive using current generation supercomputers. This is unfortunate
because, away from the caustics, characteristic evolution offers myriad
computational and geometrical advantages.

\begin{figure}[hptb]
  \def\epsfsize#1#2{0.65#1}
  \centerline{\epsfbox{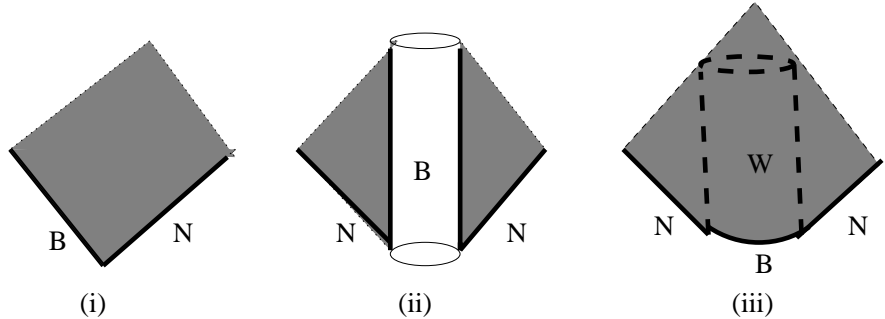}}
  \caption{\it The three applications of characteristic evolution
    with data given on an initial null hypersurface N and boundary
    B. The shaded regions indicate the corresponding domains of
    dependence.}
  \label{fig:civp}
\end{figure}

As a result, at least in the near future, fully three-dimensional computational
applications of characteristic evolution are likely to be restricted to some
mixed form, in which data is prescribed on a non-singular but incomplete
initial null hypersurface N and on a second boundary hypersurface B, which
together with the initial null hypersurface determine a nontrivial domain of
dependence. The hypersurface B may be either (i) null, (ii) timelike or
(iii) spacelike, as schematically depicted in Figure~\ref{fig:civp}. The first
two possibilities give rise to (i) the double null problem and (ii) the
nullcone-worldtube problem. Possibility (iii) has more than one interpretation.
It may be regarded as a Cauchy initial boundary value problem where the outer
boundary is null. An alternative interpretation is the Cauchy-characteristic
matching (CCM) problem, in which the Cauchy and characteristic evolutions are
matched transparently across a worldtube W, as indicated in
Figure~\ref{fig:civp}.

In CCM, it is possible to choose the matching interface between the Cauchy and
characteristic regions to be a null hypersurface, but it is more practical to
match across a timelike worldtube. CCM combines the advantages of
characteristic evolution in treating the outer radiation zone in spherical
coordinates which are naturally adapted to the topology of the worldtube with
the advantages of Cauchy evolution in treating the inner region in Cartesian
coordinates, where spherical coordinates would break down.

In this review, we trace the development of characteristic algorithms from
model 1D problems to a 2D axisymmetric code which computes the gravitational
radiation from the oscillation and gravitational collapse of a relativistic
star and to a 3D code designed to calculate the waveform emitted in the merger
to ringdown phase of a binary black hole. And we trace the development of CCM
from early feasibility studies to successful implementation in the linear
regime and through current attempts to treat the binary black hole problem.

This material includes several notable developments since my last review. Most
important for future progress have been two Ph.D.\ theses based upon
characteristic evolution codes. Florian Siebel's thesis work~\cite{Siebel}, at
the Technische Universit{\" {a}}t M{\" {u}}nchen, integrates an axisymmetric
characteristic gravitational code with a high resolution shock capturing code
for relativistic hydrodynamics. This coupled general relativistic code has been
thoroughly tested and has yielded state-of-the-art results for the
gravitational waves produced by the oscillation and collapse of a relativistic
star (see Sections~\ref{sec:shydro} and~\ref{sec:ahydro}). In Yosef Zlochower's
thesis work~\cite{Zlochower}, at the University of Pittsburgh, the
gravitational waves generated from the post-merger phase of a binary black
black hole is computed using a fully nonlinear three-dimensional characteristic
code~\cite{zlochmode} (see Section~\ref{sec:mode}). He shows how the
characteristic code can be employed to investigate the nonlinear mode coupling
in the response of a black hole to the infall of gravitational waves.

A further notable achievement has been the successful application of CCM to the
linearized matching problem between a 3D characteristic code and a 3D Cauchy
code based upon harmonic coordinates~\cite{harm} (see Section~\ref{sec:linccm}).
Here the linearized Cauchy code satisfies a well-posed initial-boundary value
problem, which seems to be a critical missing ingredient in previous
attempts at CCM in general relativity.

The problem of computing the evolution of a self-gravitating object, such as a
neutron star, in close orbit about a black hole is of clear importance to the
new gravitational wave detectors. The interaction with the black hole could be
strong enough to produce a drastic change in the emitted waves, say by tidally
disrupting the star, so that a perturbative calculation would be inadequate.
The understanding of such nonlinear phenomena requires well behaved numerical
simulations of hydrodynamic systems satisfying Einstein's equations. Several
numerical relativity codes  for treating the problem of a neutron star near a
black hole have been developed, as described in the {\it Living Review
  in Relativity} on ``Numerical Hydrodynamics in General Relativity'' by
Font~\cite{tonirev}. Although most of these efforts concentrate on Cauchy
evolution, the characteristic approach has shown remarkable robustness in
dealing with a single black hole or relativistic star. In this vein,
state-of-the-art axisymmetric studies of the oscillation and gravitational
collapse of relativistic stars have been achieved (see Section~\ref{sec:ahydro})
and progress has been made in the 3D simulation of a body in close orbit about
a Schwarzschild black hole (see Sections~\ref{sec:3dhydro} and~\ref{sec:part}).

In previous reviews, I tried to include material on the treatment of boundaries
in the computational mathematics and fluid dynamics literature because of its
relevance to the CCM problem. The fertile growth of this subject makes this
impractical to continue. A separate {\it Living Review in Relativity}
on boundary conditions is
certainly warranted and is presently under consideration. In view of this, I
will not attempt to keep this subject up to date except for material of direct
relevance to CCM, although I will for now retain the past material.

Animations and other material from these studies can be viewed at the
web sites of the University of Canberra~\cite{canberra}, Louisiana
State University~\cite{lsu}, Pittsburgh University~\cite{pitt}, and
Pittsburgh Supercomputing Center~\cite{psc}.

\newpage


\section{The Characteristic Initial Value Problem}

Characteristics have traditionally played an important role in the
analysis of hyperbolic partial differential equations. However, the use
of characteristic hypersurfaces to supply the foliation underlying an
evolution scheme has been mainly restricted to relativity. This is
perhaps natural because in curved spacetime there is no longer a
preferred Cauchy foliation provided by the Euclidean 3-spaces allowed
in Galilean or special relativity. The method of shooting along
characteristics is a standard technique in many areas of computational
physics, but evolution based upon characteristic hypersurfaces is quite
uniquely limited to relativity.

Bondi's initial use of null coordinates to describe radiation
fields~\cite{1bondi} was followed by a rapid development of other null
formalisms. These were distinguished either as metric based approaches, as
developed for axisymmetry by Bondi, Metzner and van der Burg~\cite{bondi} and
generalized to 3 dimensions by Sachs~\cite{sachs}, or as null tetrad approaches
in which the Bianchi identities appear as part of the system of equations, as
developed by Newman and Penrose~\cite{NP}.

At the outset, null formalisms were applied to construct asymptotic solutions
at null infinity by means of $1/r$ expansions. Soon afterward, Penrose devised
the conformal compactification of null infinity $\mathcal{I}$ (``scri''), thereby
reducing to geometry the asymptotic quantities describing the physical
properties of the radiation zone, most notably the Bondi mass and news
function~\cite{Penrose}. The characteristic initial value problem rapidly
became an important tool for the clarification of fundamental conceptual issues
regarding gravitational radiation and its energy content. It laid bare and
geometrised the gravitational far field.

The initial focus on asymptotic solutions clarified the kinematic properties of
radiation fields but could not supply the waveform from a specific source. It
was soon realized that instead of carrying out a $1/r$ expansion, one could
reformulate the approach in terms of the integration of ordinary differential
equations along the characteristics (null rays)~\cite{tam}. The integration
constants supplied on some inner boundary then determined the specific
waveforms obtained at infinity. In the double-null initial value problem of
Sachs~\cite{sachsdn}, the integration constants are supplied at the
intersection of outgoing and ingoing null hypersurfaces. In the
worldtube-nullcone formalism, the sources were represented by integration
constants on a timelike worldtube~\cite{tam}. These early formalisms have gone
through much subsequent revamping. Some have been reformulated to fit the
changing styles of modern differential geometry. Some have been reformulated in
preparation for implementation as computational algorithms. The articles
in~\cite{southam} give a representative sample of formalisms. Rather than
including a review of the extensive literature on characteristic formalisms in
general relativity, I concentrate here on those approaches which have been
implemented as computational evolution schemes. The existence and uniqueness of
solutions to the associated boundary value problems, which has obvious
relevance to the success of numerical simulations, is treated in a separate
{\it Living Review in Relativity} on ``Theorems on Existence and
Global Dynamics for the Einstein Equations'' by Rendall~\cite{rendall}.

All characteristic evolution schemes share the same skeletal form. The
fundamental ingredient is a foliation by null hypersurfaces
$ u = \mathrm{const.} $
which are generated by a two-dimensional set of null rays, labeled by
coordinates $x^A$, with a coordinate $\lambda$ varying along the rays.
In $(u,\lambda,x^A)$ null coordinates, the main set of Einstein
equations take the schematic form
\begin{equation}
  F_{,\lambda} = H_F [F,G]
\end{equation}
and
\begin{equation}
  G_{,u\lambda} = H_G [F,G,G_{,u}].
\end{equation}
Here $F$ represents a set of hypersurface variables, $G$ a set of evolution
variables, and $H_F$ and $H_G$ are nonlinear hypersurface operators, i.e.\ they
operate locally on the values of $F$, $G$ and $G_{,u}$ intrinsic to a single
null hypersurface. In the Bondi formalism, these hypersurface equations have a
hierarchical structure in which the members of the set $F$ can be integrated in
turn in terms of the characteristic data for the evolution variables and prior
members of the hierarchy. In addition to these main equations, there is a
subset of four Einstein equations which are satisfied by virtue of the Bianchi
identities, provided that they are satisfied on a hypersurface transverse to
the characteristics. These equations have the physical interpretation as
conservation laws. Mathematically they are analogous to the constraint
equations of the canonical formalism. But they are not necessarily elliptic,
since they may be intrinsic to null or timelike hypersurfaces, rather than
spacelike Cauchy hypersurfaces.

Computational implementation of characteristic evolution may be based upon
different versions of the formalism (i.e.\ metric or tetrad) and different
versions of the initial value problem (i.e.\ double null or worldtube-nullcone).
The performance and computational requirements of the resulting evolution codes
can vary drastically. However, most characteristic evolution codes share certain
common advantages:
\begin{itemize}
\item The initial data is free. There are no elliptic constraints on
  the data, which eliminates the need for time consuming iterative
  constraint solvers and artificial boundary conditions. This
  flexibility and control in prescribing initial data has the
  trade-off of limited experience with prescribing physically
  realistic characteristic initial data.
\item The coordinates are very``rigid'', i.e.\ there is very little
  remaining gauge freedom.
\item The constraints satisfy ordinary differential equations along
  the characteristics which force any constraint violation to fall off
  asymptotically as $1/r^2$.
\item No second time derivatives appear so that the number of basic
  variables is at most half the number for the corresponding version
  of the Cauchy problem.
\item The main Einstein equations form a system of coupled ordinary
  differential equations with respect to the parameter $\lambda$
  varying along the characteristics. This allows construction of an
  evolution algorithm in terms of a simple march along the
  characteristics.
\item In problems with isolated sources, the radiation zone can be
  compactified into a finite grid boundary with the metric rescaled by
  $1/r^2$ as an implementation of Penrose's conformal method. Because
  the Penrose boundary is a null hypersurface, no extraneous outgoing
  radiation condition or other artificial boundary condition is
  required. The analogous treatment in the Cauchy problem requires the
  use of hyperboloidal spacelike hypersurfaces asymptoting to null
  infinity~\cite{Fhprbloid}. For reviews of the hyperboloidal approach
  and its status in treating the associated three-dimensional
  computational problem, see~\cite{Hhprbloid,joerg}.
\item The grid domain is exactly the region in which waves propagate,
  which is ideally efficient for radiation studies. Since each null
  hypersurface of the foliation extends to infinity, the radiation is
  calculated immediately (in retarded time).
\item In black hole spacetimes, a large redshift at null infinity
  relative to internal sources is an indication of the formation of an
  event horizon and can be used to limit the evolution to the exterior
  region of spacetime. While this can be disadvantageous for late time
  accuracy, it allows the possibility of identifying the event horizon
  ``on the fly'', as opposed to Cauchy evolution where the event
  horizon can only be located after the evolution has been completed.
\end{itemize}
Perhaps most important from a practical view, characteristic evolution
codes have shown remarkably robust stability and were the first to
carry out long term evolutions of moving black holes~\cite{wobb}.

Characteristic schemes also share as a common disadvantage the necessity either
to deal with caustics or to avoid them altogether. The scheme to tackle the
caustics head on by including their development as part of the evolution is
perhaps a great idea still ahead of its time but one that should not be
forgotten. There are only a handful of structurally stable caustics, and they
have well known algebraic properties. This makes it possible to model their
singular structure in terms of Pad\' e approximants. The structural stability
of the singularities should in principle make this possible, and algorithms to
evolve the elementary caustics have been proposed~\cite{cstew,padstew}. In the
axisymmetric case, cusps and folds are the only structurally stable caustics,
and they have already been identified in the horizon formation occurring in
simulations of head-on collisions of black holes and in the temporarily
toroidal horizons occurring in collapse of rotating matter~\cite{sci,torus}. In
a generic binary black hole horizon, where axisymmetry is broken, there is a
closed curve of cusps which bounds the two-dimensional region on the horizon
where the black holes initially form and merge~\cite{ndata,asym}.

\newpage


\section{Prototype Characteristic Evolution Codes}

Limited computer power, as well as the instabilities arising from
non-hyperbolic formulations of Einstein's equations, necessitated that the
early code development in general relativity be restricted to spacetimes with
symmetry. Characteristic codes were first developed for spacetimes with
spherical symmetry. The techniques for relativistic fields which propagate on
null characteristics are similar to the gravitational case. Such fields
are included in this section. We postpone treatment of relativistic fluids,
whose characteristics are timelike, until Section~\ref{sec:grace}.


\subsection{\{1\,+\,1\}-dimensional codes}
\label{sec:1d}

It is often said that the solution of the general ordinary differential
equation is essentially known, in light of the success of computational
algorithms and present day computing power. Perhaps this is an
overstatement because investigating singular behavior is still an art. But,
in this spirit, it is fair to say that the general system of hyperbolic
partial differential equations in one spatial dimension seems to be a
solved problem in general relativity. Codes have been successful in
revealing important new phenomena underlying singularity formation in
cosmology~\cite{berger} and in dealing with unstable spacetimes to
discover critical phenomena~\cite{gundlach}. As described below, characteristic
evolution has contributed to a rich variety of such results.

One of the earliest characteristic evolution codes, constructed by Corkill
and Stewart~\cite{cstew,bonn}, treated spacetimes with two Killing vectors
using a grid  based upon double null coordinates, with the null
hypersurfaces intersecting in the surfaces spanned by the Killing vectors.
They simulated colliding plane waves and evolved the
Khan--Penrose~\cite{khan} collision of impulsive ($\delta$-function
curvature) plane waves to within a few numerical zones from the final
singularity, with extremely close agreement with the analytic results.
Their simulations of collisions with more general waveforms, for which
exact solutions are not known, provided input to the understanding of
singularity formation which was unforeseen in the analytic treatments of
this problem.

Many \{1\,+\,1\}-dimensional characteristic codes have been developed for spherically
symmetric systems. Here matter must be included in order to make the system
non-Schwarzschild. Initially the characteristic evolution of matter was
restricted to simple cases, such as massless Klein--Gordon fields, which allowed
simulation of gravitational collapse and radiation effects in the simple
context of spherical symmetry. Now, characteristic evolution of matter is
progressing to more complicated systems. Its application to hydrodynamics
has made significant contributions to general relativistic
astrophysics, as reviewed in Section~\ref{sec:grace}.

The synergy between analytic and computational approaches has already led to
dramatic results in the massless Klein--Gordon case. On the analytic side,
working in a characteristic initial value formulation based upon outgoing null
cones, Christodoulou made a penetrating study of the spherically symmetric
problem~\cite{X21987,X1991,X1993,X1994,X1999,X21999}. In a suitable function
space, he showed the existence of an open ball about Minkowski space data whose
evolution is a complete regular spacetime; he showed that an evolution with a
nonzero final Bondi mass forms a black hole; he proved a version of cosmic
censorship for generic data; and he established the existence of naked
singularities for non-generic data. What this analytic tour-de-force did not
reveal was the remarkable critical behavior in the transition to the black
hole regime, which was discovered by Choptuik~\cite{chsouth,choptprl} by
computational simulation based upon Cauchy evolution. This phenomenon has now
been understood in terms of the methods of renormalization group theory and
intermediate asymptotics, and has spawned a new subfield in general relativity,
which is covered in the {\it Living Review in Relativity} on
``Critical Phenomena in Gravitational Collapse'' by Gundlach~\cite{gundlach}.

The characteristic evolution algorithm for the spherically symmetric
Einstein--Klein--Gordon problem provides a simple illustration of the
techniques used in the general case. It centers about the evolution scheme
for the scalar field, which constitutes the only dynamical field. Given the
scalar field, all gravitational quantities can be determined by integration
along the characteristics of the null foliation. This is a coupled problem,
since the scalar wave equation involves the curved space metric. It
illustrates how null algorithms lead to a hierarchy of equations which can
be integrated along the characteristics to effectively decouple the
hypersurface and dynamical variables.

In a Bondi coordinate system based upon outgoing null hypersurfaces
$u = \mathrm{const.}$ and a surface area coordinate $r$, the metric is
\begin{equation}
  ds^2 = -e^{2\beta} du \left( \frac{V}{r} du + 2 \, dr \right) +
  r^2 \left( d\theta^2 + \sin^2 \theta \, d\phi^2 \right).
  \label{eq:metric}
\end{equation}
Smoothness at $r=0$ allows imposition of the coordinate conditions
\begin{equation}
  \begin{array}{rcl}
    V(u,r) &=& r + \mathcal{O} (r^3) \\ [0.5 em]
    \beta(u,r) &=& \mathcal{O} (r^2).
  \end{array}
  \label{eq:bc}
\end{equation}
The field equations consist of the curved space wave equation $\Box \Phi = 0$
for the scalar field and two hypersurface equations for the metric
functions:
\begin{eqnarray}
  \beta_{,r} &=& 2 \pi r (\Phi_{,r})^2,
  \label{eq:sbeta}
  \\
  V_{,r} &=& e^{2 \beta}.
  \label{eq:sv}
\end{eqnarray}%
The wave equation can be expressed in the form
\begin{equation}
  \Box^{(2)} g - \left( \frac{V}{r} \right)_{\!\!,r}
  \frac{e^{-2 \beta} g}{r} = 0,
  \label{eq:hatwave}
\end{equation}
where $g=r\Phi$ and $\Box ^{(2)}$ is the D'Alembertian associated with
the two-dimensional submanifold spanned by the ingoing and outgoing
null geodesics. Initial null data for evolution consists of
$\Phi(u_0,r)$ at initial retarded time $u_0$.

\begin{figure}[hptb]
  \def\epsfsize#1#2{0.55#1}
  \centerline{\epsfbox{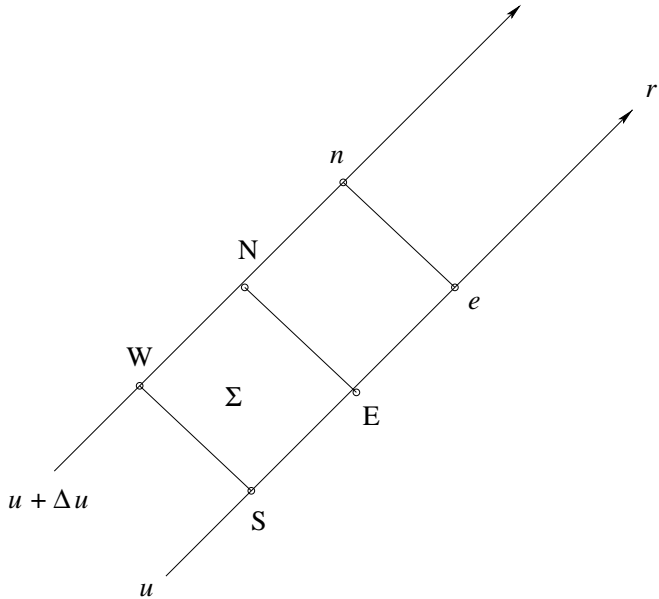}}
  \caption{\it The null parallelogram. After computing the field at
    point $N$, the algorithm marches the computation to $\mathcal{I}^+$
    by shifting the corners by $N\rightarrow n$, $E\rightarrow e$,
    $S\rightarrow E$, $W\rightarrow N$.}
  \label{fig:nsew}
\end{figure}

Because any two-dimensional geometry is conformally flat, the surface
integral of $\Box^{(2)} g$ over a null parallelogram $\Sigma$ gives
exactly the same result as in a flat 2-space, and leads to an integral
identity upon which a simple evolution algorithm can be
based~\cite{scasym}. Let the vertices of the null parallelogram be
labeled by $N$, $E$, $S$, and $W$ corresponding, respectively, to their
relative locations (North, East, South, and West) in the 2-space, as
shown in Figure~\ref{fig:nsew}. Upon integration of Equation~(\ref{eq:hatwave}),
curvature introduces an integral correction to the flat space
null parallelogram relation between the values of $g$ at the vertices:
\begin{equation}
  g_N - g_W - g_E + g_S = - \frac{1}{2} \int_\Sigma du \, dr
  \left( \frac{V}{r} \right)_{\!\!,r} \frac{g}{r}.
  \label{eq:integral}
\end{equation}

This identity, in one form or another, lies behind all of the null
evolution algorithms that have been applied to this system. The prime
distinction between the different algorithms is whether they are based
upon double null coordinates or Bondi coordinates as in
Equation~(\ref{eq:metric}). When a double null coordinate system is
adopted, the points  $N$, $E$, $S$, and $W$ can be located in each
computational cell at grid points, so that evaluation of the left hand
side of Equation~(\ref{eq:integral}) requires no interpolation. As a
result, in flat space, where the right hand side of
Equation~(\ref{eq:integral}) vanishes, it is possible to formulate an
exact evolution algorithm. In curved space, of course, there is a truncation
error arising from the approximation of the integral, e.g., by evaluating the
integrand at the center of $\Sigma$.

The identity~(\ref{eq:integral}) gives rise to the following explicit marching
algorithm, indicated in Figure~\ref{fig:nsew}. Let the null parallelogram lie at
some fixed $\theta$ and $\phi$ and span adjacent retarded time levels $u_0$ and
$u_0+\Delta u$. Imagine for now that the points $N$, $E$, $S$, and $W$ lie on
the spatial grid, with $r_N-r_W=r_E-r_S=\Delta r$. If $g$ has been determined
on the entire initial cone $u_0$, which contains the points $E$ and $S$, and
$g$ has been determined radially outward from the origin to the point $W$ on
the next cone $u_0+\Delta u$, then Equation~(\ref{eq:integral}) determines $g$ at
the next radial grid point $N$ in terms of an integral over $\Sigma$. The
integrand can be approximated to second order, i.e.\ to $\mathcal{O} (\Delta r \Delta u)$,
by evaluating it at the center of $\Sigma$. To this same accuracy, the value of
$g$ at the center equals its average between the points $E$ and $W$, at which
$g$ has already been determined. Similarly, the value of $(V/r)_{,r}$ at the
center of $\Sigma$ can be approximated to second order in terms of values of
$V$ at points where it can be determined by integrating the hypersurface
equations~(\ref{eq:sbeta}, \ref{eq:sv}) radially outward from $r=0$.

After carrying out this procedure to evaluate $g$ at the point $N$, the
procedure can then be iterated to determine $g$ at the next radially outward
grid point on the $u_0+\Delta u$ level, i.e.\ point $n$ in Figure~\ref{fig:nsew}.
Upon completing this radial march to null infinity, in terms of a compactified
radial coordinate such as $x=r/(1+r)$, the field $g$ is then evaluated on the
next null cone at $u_0+2\Delta u$, beginning at the vertex where smoothness
gives the startup condition that $g(u,0)=0$.

In the compactified Bondi formalism, the vertices $N$, $E$, $S$, and $W$
of the null parallelogram $\Sigma$ cannot be chosen to lie exactly on
the grid because, even in Minkowski space, the velocity of light in
terms of a compactified radial coordinate $x$ is not constant. As a
consequence, the fields $g$, $\beta$, and $V$ at the vertices of
$\Sigma$ are approximated to second order accuracy by interpolating
between grid points. However, cancellations arise between these four
interpolations so that Equation~(\ref{eq:integral}) is satisfied to fourth
order accuracy. The net result is that the finite difference version
of Equation~(\ref{eq:integral}) steps $g$ radially outward one zone with an
error of fourth order in grid size, $\mathcal{O} ((\Delta u)^2 (\Delta x)^2)$. In
addition, the smoothness conditions~(\ref{eq:bc}) can be incorporated
into the startup for the numerical integrations for $V$ and $\beta$ to
insure no loss of accuracy in starting up the march at $r=0$. The
resulting global error in $g$, after evolving a finite retarded time,
is then $\mathcal{O} (\Delta u\Delta x)$, after compounding errors from $1/(\Delta
u\Delta x)$ number of zones.

When implemented on a grid based upon the $(u,r)$ coordinates, the stability of
this algorithm is subject to a Courant--Friedrichs--Lewy (CFL) condition
requiring that the physical domain of dependence be contained in the numerical
domain of dependence. In the spherically symmetric case, this condition
requires that the ratio of the time step to radial step be limited by
$(V/r)\Delta u \le 2\Delta r$, where $\Delta r=\Delta[x/(1-x)]$. This condition
can be built into the code using the value $V/r=e^{2H}$, corresponding to the
maximum of $V/r$ at $\cal I^+$. The strongest restriction on the time step then
arises just before the formation of a horizon, where $V/r\rightarrow \infty$ at
$\cal I^+$. This infinite redshift provides a mechanism for locating the true
event horizon ``on the fly'' and restricting the evolution to the exterior
spacetime. Points near $\cal I^+$ must be dropped in order to evolve across
the horizon due to the lack of a nonsingular compactified version of future
time infinity  $I^+$.

The situation is quite different in a double null coordinate system, in which
the vertices of the null parallelogram can be placed exactly on grid points so
that the CFL condition is automatically satisfied. A characteristic code based
upon double null coordinates was developed by Goldwirth and Piran in a study of
cosmic censorship~\cite{goldw} based upon the spherically symmetric
gravitational collapse of a massless scalar field. Their early study lacked the
sensitivity of adaptive mesh refinement (AMR) which later enabled Choptuik to
discover the critical phenomena appearing in this problem. Subsequent work by
Marsa and Choptuik~\cite{mc} combined the use of the null related ingoing
Eddington--Finklestein coordinates with Unruh's strategy of singularity
excision to construct a 1D code that ``runs forever''. Later,
Garfinkle~\cite{garf1} constructed an improved version of  the Goldwirth--Piran
double null code which was able to simulate critical phenomena without using
adaptive mesh refinement. In this treatment, as the evolution proceeds on one
outgoing null cone to the next, the grid points follow the ingoing null cones
and must be dropped as they cross the origin at $r=0$. However, after half the
grid points are lost they are  then ``recycled'' at new positions midway
between the remaining grid points. This technique is crucial for resolving the
critical phenomena associated with an $r\rightarrow  0$ size horizon. An
extension of the code~\cite{garf2} was later used to verify that scalar field
collapse in six dimensions continues to display critical phenomena.

Hamad\'e and Stewart~\cite{hamst} also applied a double null code to study
critical phenomena. In order to obtain the accuracy necessary to confirm
Choptuik's results they developed the first example of a characteristic grid
with AMR. They did this with both the standard
Berger and Oliger algorithm and their own simplified version, with both
versions giving indistinguishable results. Their simulations of critical
collapse of a massless scalar field agreed with Choptuik's values for the
universal parameters governing mass scaling and displayed the echoing
associated with discrete self-similarity. Hamad\'e, Horne, and
Stewart~\cite{hamhst} extended this study to the spherical collapse of an
axion/dilaton system and found in this case that self-similarity was a
continuous symmetry of the critical solution.

Brady, Chambers, and Goncalves~\cite{BrChGo} used Garfinkle's~\cite{garf1}
double null algorithm to investigate  the effect of a \emph{massive} scalar
field on critical phenomena. The introduction of a mass term in the scalar wave
equation introduces a scale to the problem, which suggests that the critical
point behavior might differ from the massless case. They found that there are
two different regimes depending on the ratio of the Compton wavelength $1/m$ of
the scalar mass to the radial size $\lambda$ of the scalar pulse used to induce
collapse. When $\lambda m << 1$, the critical solution is the one found by
Choptuik in the $m=0$ case, corresponding to a type~II phase transition.
However, when $\lambda m >> 1$, the critical solution is an unstable soliton
star (see~\cite{SeidelSuen}), corresponding to a type~I phase transition where
black hole formation turns on at a finite mass.

A code based upon Bondi coordinates, developed by Husa and his
collaborators~\cite{huscrit}, has been successfully applied to spherically
symmetric critical collapse of a nonlinear $\sigma$-model coupled to gravity.
Critical phenomena cannot be resolved on a static grid based upon the Bondi
$r$-coordinate. Instead, the numerical techniques of Garfinkle were adopted by
using a dynamic grid following the ingoing null rays and by recycling radial
grid points. They studied how coupling to gravity affects the critical behavior
previously observed by Bizo\'n~\cite{bizon} and others in the Minkowski
space version of the model. For a wide range of the coupling constant, they
observe discrete self-similarity and typical mass scaling near the critical
solution. The code is shown to be second order accurate and to give second
order convergence for the value of the critical parameter.

The first characteristic code in Bondi coordinates for the self-gravitating
scalar wave problem was constructed by G\'omez and Winicour~\cite{scasym}. They
introduced a numerical compactification of $\mathcal{I}^+$ for the purpose of
studying  effects of self-gravity on the scalar radiation, particularly in the
high amplitude limit of the rescaling $\Phi\rightarrow a\Phi$. As $a
\rightarrow\infty$, the red shift creates an effective boundary layer at
$\mathcal{I}^+$ which causes the Bondi mass $M_\mathrm{B}$ and the scalar field monopole
moment $Q$ to be related by $M_\mathrm{B}\sim \pi |Q|/\sqrt{2}$, rather than the
quadratic relation of the weak field limit~\cite{scasym}. This could also be
established analytically so that the high amplitude limit provided a check on
the code's ability to handle strongly nonlinear fields. In the small amplitude
case, this work \emph{incorrectly} reported that the radiation tails from black
hole formation had an exponential decay characteristic of quasinormal modes
rather than the polynomial $1/t$ or $1/t^2$ falloff expected from
Price's~\cite{price} work on perturbations of Schwarzschild black
holes. In hindsight,
the error here was not having confidence to run the code sufficiently long to
see the proper late time behavior.

Gundlach, Price, and Pullin~\cite{gpp1,gpp2} subsequently reexamined the
issue of power law tails using a double null code similar to that developed by
Goldwirth and Piran. Their numerical simulations verified the existence of
power law tails in the full nonlinear case, thus establishing consistency with
analytic perturbative theory. They also found normal mode ringing at
intermediate time, which provided reassuring consistency with perturbation
theory and showed that there is a region of spacetime where the results of
linearized theory are remarkably reliable even though highly nonlinear behavior
is taking place elsewhere. These results have led to a methodology that has
application beyond the confines of spherically symmetric problems, most notably
in the ``close approximation'' for the binary black hole problem~\cite{jorge}.
Power law tails and quasinormal ringing have also been confirmed using Cauchy
evolution~\cite{mc}.

The study of the radiation tail decay of a scalar field was subsequently
extended by G\'omez, Schmidt, and Winicour~\cite{gsw} using a characteristic
code. They showed that the Newman--Penrose constant~\cite{conserved} for the
scalar field determines the exponent of the power law (and not the static
monopole moment as often stated). When this constant is non-zero, the tail
decays as $1/t$ on $\mathcal{I}^+$, as opposed to the $1/t^2$ decay for the
vanishing case. (They also found $t^{-n}\log t$ corrections, in addition to
the exponentially decaying contributions of the quasinormal modes.)  This code
was also used to study the instability of a topological kink in the
configuration of the scalar field~\cite{kink}. The kink instability provides
the simplest example of the turning point instability~\cite{ipser,sork} which
underlies gravitational collapse of static equilibria.

Brady and Smith~\cite{bradysmith} have demonstrated that characteristic
evolution is especially well adapted to explore properties of Cauchy horizons.
They examined the stability of the Reissner--Nordstr\"om Cauchy horizon using an
Einstein--Klein--Gordon code based upon advanced Bondi coordinates $(v,r)$ (where
the hypersurfaces $v=const$ are ingoing null hypersurfaces). They study the
effect of a spherically symmetric scalar pulse on the spacetime structure as it
propagates across the event horizon. Their numerical methods are patterned
after the work of Goldwirth and Piran~\cite{goldw}, with modifications of the
radial grid structure that allow deep penetration inside the black hole. In
accord with expectations from analytic studies, they find that the pulse first
induces a weak null singularity on the Cauchy horizon, which then
leads to a crushing spacelike singularity as $r\rightarrow 0$. The null
singularity is weak in the sense that an infalling observer experiences a
finite tidal force, although the Newman--Penrose Weyl component $\Psi_2$
diverges, a phenomenon known as mass inflation~\cite{poisson}. These results
confirm the earlier result of Gnedin and Gnedin~\cite{gnedin} that a central
spacelike singularity would be created by the interaction of a charged black
hole with a scalar field, in accord with a physical argument by
Penrose~\cite{penrcoscen} that a small perturbation undergoes an infinite
redshift as it approaches the Cauchy horizon.

Burko~\cite{burko} has confirmed and extended these results, using a code based
upon double null coordinates which was developed with Ori~\cite{burkori} in a
study of tail decay. He found that in the early stages the
perturbation of the Cauchy horizon is weak and in agreement with the behavior
calculated by perturbation theory.

Brady, Chambers, Krivan, and Laguna~\cite{BrChKrLa} have found interesting
effects of a non-zero cosmological constant $\Lambda$ on tail decay by using a
characteristic Einstein--Klein--Gordon code to study the effect of a massless
scalar pulse on Schwarzschild--de Sitter and Reissner--Nordstr\"om--de Sitter
spacetimes. First, by constructing a linearized scalar evolution code, they
show that scalar test fields with $\ell\ne 0$ have exponentially decaying
tails, in contrast to the standard power law tails for asymptotically flat
spacetimes. Rather than decaying, the monopole mode asymptotes at late time to
a constant, which scales linearly with $\Lambda$, in contrast to the standard
no-hair result. This unusual behavior for the $\ell=0$ case was then
independently confirmed with a nonlinear spherical characteristic code.

Using a combination of numerical and analytic techniques based upon null
coordinates, Hod and Piran have made an extensive series of investigations of
the spherically symmetric charged Einstein--Klein--Gordon system dealing with the
effect of charge on critical gravitational collapse~\cite{HodPir1} and the late
time tail decay of a charged scalar field on a Reissner--Nordstr\"om black
hole~\cite{HodPir2,HodPir2.5,HodPir3,HodPir4}. These studies culminated in a
full nonlinear investigation of horizon formation by the collapse of a charged
massless scalar pulse~\cite{HodPir5}. They track the formation of an apparent
horizon which is followed by a weakly singular Cauchy horizon which develops a
strong spacelike singularity at $r=0$. This is in complete accord with prior
perturbative results and nonlinear simulations involving a pre-existing black
hole. Oren and Piran~\cite{OrenPir} increased the late time accuracy of this
study by incorporating an adaptive grid for the retarded time coordinate $u$,
with a refinement criterion to maintain $\Delta r/r = \mathrm{const}$. The accuracy of
this scheme is confirmed through convergence tests as well as charge and
constraint conservation. They were able to observe the physical mechanism which
prohibits black hole formation with charge to mass ration $Q/M>1$.
Electrostatic repulsion of the outer parts of the scalar pulse increases
relative to the gravitational attraction and causes the outer portion of the
charge to disperse to larger radii before the black hole is formed. Inside the
black hole, they confirm the formation of a weakly singular Cauchy horizon
which turns into a strong spacelike singularity, in accord with other studies.

Hod extended this combined numerical-analytical double null approach to
investigate higher order corrections to the dominant power law
tail~\cite{hod2}, as well as corrections due to a general spherically symmetric
scattering potential~\cite{hod1} and due to a time dependent
potential~\cite{hod3}. He found $(\log t)/t$ modifications to the leading order
tail behavior for a Schwarzschild black hole, in accord with earlier results of
G{\'{o}}mez et al.~\cite{gsw}. These modifications fall off at a slow rate
so that a very long numerical evolution ($t\approx 3000 \, M$)is
necessary to cleanly identify the leading order power law decay.

The foregoing numerical-analytical work based upon characteristic evolution has
contributed to a very comprehensive classical treatment of spherically
symmetric gravitational collapse. Sorkin and Piran~\cite{SorkPir} have
investigated the question of quantum  corrections due to pair creation on the
gravitational collapse of a charged scalar field. For observers outside the
black hole, several analytic studies have indicated that such pair-production
can rapidly diminish the charge of the black hole. Sorkin and Piran apply the
same double-null characteristic code used in studying the classical
problem~\cite{HodPir5} to evolve across the event horizon and observe the
quantum effects on the Cauchy horizon. The quantum electrodynamic effects are
modeled in a rudimentary way by a nonlinear dielectric $\epsilon$ constant that
limits the electric field to the critical value necessary for pair creation.
The back-reaction of the pairs on the stress-energy and the electric current
are ignored. They found that quantum effects leave the classical picture of the
Cauchy horizon qualitatively intact but that they shorten its ``lifetime'' by
hastening the conversion of the weak null singularity into a strong spacelike
singularity.

The Southampton group has constructed a \{1\,+\,1\}-dimensional
characteristic code for spacetimes with cylindrical
symmetry~\cite{cylinder1,cylinder2}. The original motivation was to use it
as the exterior characteristic code in a test case of CCM (see
Section~\ref{sec:cylmatch} for the
application to matching). Subsequently, Sperhake, Sj\" odin, and
Vickers~\cite{vick1,vick2} modified the code into a global
characteristic version for the purpose of studying cosmic strings,
represented by massive scalar and vector fields coupled to gravity. Using a
Geroch decomposition~\cite{gerdec} with respect to the translational
Killing vector, they reduced the global problem to a \{2\,+\,1\}-dimensional
asymptotically flat spacetime, so that $\mathcal{I}^+$ can be compactified and
included in the numerical grid. Rather than the explicit scheme used in
CCM, the new version employs an implicit, second order in space and time,
Crank--Nicholson evolution scheme. The code showed long term stability and
second order convergence in vacuum tests based upon exact Weber--Wheeler
waves~\cite{wweb} and Xanthopoulos' rotating solution~\cite{xanth}, and in
tests of wave scattering by a string. The results show damped ringing of
the string after an incoming Weber--Wheeler pulse has excited it and then
scattered to $\mathcal{I}^+$. The ringing frequencies are independent of
the details of the pulse but are inversely proportional to the masses of
the scalar and vector fields.


\subsubsection{Adaptive mesh refinement}

The goal of computing waveforms from relativistic binaries, such as a neutron
star spiraling into a black hole, requires more than a stable convergent code.
It is a delicate task to extract a waveform in a spacetime in which there are
multiple length scales: the size of the black hole, the size of the star, the
wavelength of the radiation. It is commonly agreed that some form of mesh
refinement is essential to attack this problem. Mesh refinement was first
applied in characteristic evolution to solve specific spherically symmetric
problems regarding critical phenomena and singularity
structure~\cite{garf1,hamst,burko}.

Pretorius and Lehner~\cite{pretlehn} have presented a general approach for
AMR to a generic characteristic code.
Although the method is designed to treat 3D simulations, the implementation has
so far been restricted to the Einstein--Klein--Gordon system in spherical
symmetry. The 3D approach is modeled after the Berger and Oliger AMR algorithm
for hyperbolic Cauchy problems, which is reformulated in terms of null
coordinates. The resulting characteristic AMR algorithm can be applied to any
unigrid characteristic code and is amenable to parallelization. They applied it
to the problem of a massive Klein--Gordon field propagating outward from a black
hole. The non-zero rest mass restricts the Klein--Gordon field from
propagating to infinity. Instead it diffuses into higher frequency components
which Pretorius and Lehner show can be resolved using AMR but not with a
comparison unigrid code.


\subsection{\{2\,+\,1\}-dimensional codes}

One-dimensional characteristic codes enjoy a very special simplicity
due to the two preferred sets (ingoing and outgoing) of characteristic
null hypersurfaces. This eliminates a source of gauge freedom that
otherwise exists in either two- or three-dimensional characteristic
codes. However, the manner in which the characteristics of a hyperbolic
system determine domains of dependence and lead to propagation
equations for shock waves is the same as in the one-dimensional
case. This makes it desirable for the purpose of numerical evolution to
enforce propagation along characteristics as extensively as possible.
In basing a Cauchy algorithm upon shooting along characteristics, the
infinity of characteristic rays (technically, \emph{bicharacteristics})
at each point leads to an arbitrariness which, for a practical
numerical scheme, makes it necessary either to average the propagation
equations over the sphere of characteristic directions or to select out
some preferred subset of propagation equations. The latter
approach was successfully applied by Butler~\cite{butler} to the
Cauchy evolution of two-dimensional fluid flow, but there seems to have
been very little follow-up along these lines.

The formal ideas behind the construction of two- or three-dimensional
characteristic codes are similar, although there are various technical options
for treating the angular coordinates which label the null rays. Historically,
most characteristic work graduated first from 1D to 2D because of the available
computing power.


\subsection{The Bondi problem}

The first characteristic code based upon the original Bondi equations for a
twist-free axisymmetric spacetime was constructed by Isaacson, Welling, and
Winicour~\cite{isaac} at Pittsburgh. The spacetime was foliated by a family of
null cones, complete with point vertices at which regularity conditions were
imposed. The code accurately integrated the hypersurface and evolution
equations out to compactified null infinity. This allowed studies of the Bondi
mass and radiation flux on the initial null cone, but it could not be used as a
practical evolution code because of instabilities.

These instabilities came as a rude shock and led to a retreat to the
simpler problem of axisymmetric scalar waves propagating in Minkowski
space, with the metric
\begin{equation}
  ds^2= - du^2 - 2 \, du \, dr +
  r^2 \left( d\theta^2 + \sin^2 \theta \, d\phi^2 \right)
  \label{eq:mink}
\end{equation}
in outgoing null cone coordinates. A null cone code for this problem was
constructed using an algorithm based upon Equation~(\ref{eq:integral}), with the angular
part of the flat space Laplacian replacing the curvature terms in the integrand
on the right hand side. This simple setting allowed one source of instability
to be traced to a subtle violation of the CFL condition near the vertices of
the cones. In terms of the grid spacing $\Delta x^{\alpha}$, the CFL condition
in this coordinate system takes the explicit form
\begin{equation}
  \frac{\Delta u}{\Delta r} < - 1 +
  \left[ K^2 + (K - 1)^2 - 2 K (K - 1) \cos \Delta \theta \right]^{1/2},
  \label{eq:cfl}
\end{equation}
where the coefficient $K$, which is of order 1, depends on the
particular startup procedure adopted for the outward integration. Far
from the vertex, the condition~(\ref{eq:cfl}) on the time step $\Delta
u$ is quantitatively similar to the CFL condition for a standard
Cauchy evolution algorithm in spherical coordinates. But
condition~(\ref{eq:cfl}) is strongest near the vertex of the cone
where (at the equator $\theta =\pi/2$) it implies that
\begin{equation}
  \Delta u < K \, \Delta r \, (\Delta \theta)^2.
  \label{eq:cfl0}
\end{equation}
This is in contrast to the analogous requirement
\begin{equation}
  \Delta u < K \, \Delta r \, \Delta \theta
\end{equation}
for stable Cauchy evolution near the origin of a spherical coordinate
system. The extra power of $\Delta \theta$ is the price that must be
paid near the vertex for the simplicity of a characteristic code.
Nevertheless, the enforcement of this condition allowed efficient
global simulation of axisymmetric scalar waves. Global studies of
backscattering, radiative tail decay, and solitons were carried out
for nonlinear axisymmetric waves~\cite{isaac}, but three-dimensional
simulations extending to the vertices of the cones were impractical
at the time on existing machines.

Aware now of the subtleties of the CFL condition near the vertices, the
Pittsburgh group returned to the Bondi problem, i.e.\ to evolve the Bondi
metric~\cite{bondi}
\begin{equation}
  ds^2 = \left( \frac{V}{r} e^{2 \beta} - U^2 r^2 e^{2 \gamma} \right) du^2 +
  2 e^{2 \beta} du \, dr + 2 U r^2 e^{2 \gamma} du \, d\theta -
  r^2 \left (e^{2 \gamma} d\theta^2 + e^{-2\gamma} \sin^2 \theta \, d\phi^2 \right),
  \label{eq:bmetric}
\end{equation}
by means of the three hypersurface equations
\begin{eqnarray}
  \beta_{,r} &=& \frac{1}{2} r (\gamma_{,r})^2,
  \label{eq:beta}
  \\
  \left[ r^4 e^{2 (\gamma - \beta)} U_{,r} \right]_{,r} &=&
  2 r^2 \left[ r^2  \left( \frac{\beta}{r^2} \right)_{\!\!,r\theta} -
  \frac{(\sin^2 \theta \, \gamma)_{,r\theta}}{\sin^2 \theta} +
  2 \gamma_{,r} \gamma_{,\theta} \right],
  \label{eq:U}
  \\
  V_{,r} &=& - \frac{1}{4} r^4 e^{2 (\gamma - \beta)}(U_{,r})^2 +
  \frac{(r^4 \sin \theta \, U)_{,r\theta}}{2 r^2 \sin \theta}
  \nonumber \\
  && + e^{2 (\beta - \gamma)} \left[ 1 -
  \frac{(\sin \theta \, \beta_{,\theta})_{,\theta}}{\sin \theta} +
  \gamma_{,\theta\theta} + 3 \cot \theta \, \gamma_{,\theta} -
  (\beta_{,\theta})^2 - 2 \gamma_{,\theta}
  (\gamma_{,\theta} - \beta_{,\theta}) \right], \qquad
  \label{eq:V}
\end{eqnarray}%
and the evolution equation
\begin{eqnarray}
  4 r ( r \gamma)_{,ur} & = & \left\{ 2 r \gamma_{,r} V -
  r^2 \left[ 2 \gamma_{,\theta} U + \sin \theta
  \left( \frac{U}{\sin \theta} \right)_{\!\!,\theta} \right]
  \right\}_{,r} \!\!\!\! - 2 r^{2} \frac{(\gamma_{,r} U
  \sin \theta)_{,\theta}}{\sin \theta}
  \nonumber \\
  && + \frac{1}{2} r^{4} e^{2 (\gamma - \beta)}
  (U_{,r})^2 + 2 e^{2 (\beta - \gamma)} \left[ (\beta_{,\theta})^2 +
  \sin \theta \left( \frac{\beta_{,\theta}}{\sin \theta}
  \right)_{\!\!,\theta} \right].
  \label{eq:gammaev}
\end{eqnarray}%

The beauty of the Bondi equations is that they form a clean hierarchy. Given
$\gamma$ on an initial null hypersurface, the equations can be integrated
radially to determine $\beta$, $U$, $V$, and $\gamma_{,u}$ on the hypersurface
(in that order) in terms of integration constants determined by boundary
conditions, or smoothness if extended to the vertex of a null cone. The
initial data $\gamma$ is unconstrained except by smoothness conditions. Because
$\gamma$ represents an axisymmetric spin-2 field, it must be $\mathcal{O} (\sin^2 \theta)$
near the poles of the spherical coordinates and must consist of $l\ge 2$ spin-2
multipoles.

In the computational implementation of this system by the Pittsburgh
group~\cite{papa}, the null hypersurfaces were chosen to be complete
null cones with nonsingular vertices, which (for simplicity) trace out
a geodesic worldline $r=0$. The smoothness conditions at the vertices
were formulated in local Minkowski coordinates.

The vertices of the cones were not the chief source of
difficulty. A null parallelogram marching algorithm, similar to that
used in the scalar case, gave rise to another instability that sprang up
throughout the grid. In order to reveal the source of this instability,
physical considerations suggested looking at the linearized version of
the Bondi equations, where they can be related to the wave equation.
If this relationship were sufficiently simple, then the scalar wave
algorithm could be used as a guide in stabilizing the evolution of
$\gamma$. A scheme for relating $\gamma$ to solutions $\Phi$ of the
wave equation had been formulated in the original paper by Bondi,
Metzner, and van der Burgh~\cite{bondi}. However, in that scheme, the
relationship of the scalar wave to $\gamma$ was nonlocal in the
angular directions and was not useful for the stability analysis.

A local relationship between $\gamma$ and solutions of the wave
equation was found~\cite{papa}. This provided a test bed for the null
evolution algorithm similar to the Cauchy test bed provided by
Teukolsky waves~\cite{teuk}. More critically, it allowed a simple von
Neumann linear stability analysis of the finite difference equations,
which revealed that the evolution would be unstable if the metric
quantity $U$ was evaluated on the grid. For a stable
algorithm, the grid points for $U$ must be staggered between the grid
points for $\gamma$, $\beta$, and $V$. This unexpected feature
emphasizes the value of linear stability analysis in formulating stable
finite difference approximations.

It led to an axisymmetric code~\cite{papath,papa} for the global Bondi problem
which ran stably, subject to a CFL condition, throughout the regime in which
caustics and horizons did not form. Stability in this regime was verified
experimentally by running arbitrary initial data until it radiated away to
$\mathcal{I}^+$. Also, new exact solutions as well as the linearized  null
solutions were used to perform extensive convergence tests that established
second order accuracy. The code generated a large complement of highly accurate
numerical solutions for the class of asymptotically flat, axisymmetric vacuum
spacetimes, a class for which no analytic solutions are known. All results of
numerical evolutions in this regime were consistent with the theorem of
Christodoulou and Klainerman~\cite{XKlain} that weak initial data evolve
asymptotically to Minkowski space at late time.

An additional global check on accuracy was performed  using Bondi's
formula relating mass loss to the time integral of the square of the
news function. The Bondi mass loss formula is not one of the equations
used in the evolution algorithm but follows from those equations as a
consequence of a global integration of the Bianchi identities. Thus it
not only furnishes a valuable tool for physical interpretation but it
also provides a very important calibration of numerical accuracy and
consistency.

An interesting feature of the evolution arises in regard to
compactification. By construction, the $u$-direction is timelike at
the origin where it coincides with the worldline traced out by the
vertices of the outgoing null cones. But even for weak fields, the
$u$-direction generically becomes spacelike at large distances along an
outgoing ray. Geometrically, this reflects the property that $\mathcal{I}$
is itself a null hypersurface so that all internal directions are
spacelike, except for the null generator. For a flat space time, the
$u$-direction picked out at the origin leads to a null evolution
direction at $\mathcal{I}$, but this direction becomes spacelike under a
slight deviation from spherical symmetry. Thus the evolution
generically becomes ``superluminal'' near $\mathcal{I}$. Remarkably,
this leads to no adverse numerical effects. This fortuitous property
apparently arises from the natural way that causality is built into the
marching algorithm so that no additional resort to numerical
techniques, such as ``causal differencing''~\cite{Alliance97b}, is
necessary.


\subsubsection{The conformal-null tetrad approach}

Stewart has implemented a characteristic evolution code which handles the
Bondi problem by a null tetrad, as opposed to metric, formalism~\cite{stewbm}.
The geometrical algorithm underlying the evolution scheme, as outlined
in~\cite{friedst1,friedst2}, is Friedrich's~\cite{fried}
conformal-null
description of a compactified spacetime in terms of a first order system of
partial differential equations. The variables include the metric, the
connection, and the curvature, as in a Newman--Penrose formalism, but in
addition the conformal factor (necessary for compactification of $\mathcal{I}$)
and its gradient. Without assuming any symmetry, there are more than 7 times as
many variables as in a metric based null scheme, and the corresponding
equations do not decompose into as clean a hierarchy. This disadvantage,
compared to the metric approach, is balanced by several advantages:
\begin{itemize}
\item The equations form a symmetric hyperbolic system so that
  standard theorems can be used to establish that the system is
  well-posed.
\item Standard evolution algorithms can be invoked to ensure numerical
  stability.
\item The extra variables associated with the curvature tensor are not
  completely excess baggage, since they supply essential physical
  information.
\item The regularization necessary to treat $\mathcal{I}$ is built in
  as part of the formalism so that no special numerical regularization
  techniques are necessary as in the metric case. (This last advantage
  is somewhat offset by the necessity of having to locate
  $\mathcal{I}$ by tracking the zeroes of the conformal factor.)
\end{itemize}

The code was intended to study gravitational waves from an axisymmetric star.
Since only the vacuum equations are evolved, the outgoing radiation from the
star is represented by data ($\Psi_4$ in Newman--Penrose notation) on an ingoing
null cone forming the inner boundary of the evolved domain. The inner boundary
data is supplemented by Schwarzschild data on the initial outgoing null cone,
which models an initially quiescent state of the star. This provides the
necessary data for a double-null initial value problem. The evolution would
normally break down where the ingoing null hypersurface develops caustics. But
by choosing a scenario in which a black hole is formed, it is possible to
evolve the entire region exterior to the horizon. An obvious test bed is the
Schwarzschild spacetime for which a numerically satisfactory evolution was
achieved (although convergence tests were not reported).

Physically interesting results were obtained by choosing data corresponding to
an outgoing quadrupole pulse of radiation. By increasing the initial amplitude
of the data $\Psi_4$, it was possible to evolve into a regime where the energy
loss due to radiation was large enough to drive the total Bondi mass negative.
Although such data is too grossly exaggerated to be consistent with an
astrophysically realistic source, the formation of a negative mass was an
impressive test of the robustness of the code.


\subsubsection{Axisymmetric mode coupling}
\label{sec:papamode}

Papadopoulos~\cite{papamode} has carried out an illuminating study of mode
mixing by computing the evolution of a pulse emanating outward from an
initially Schwarzschild white hole of mass $M$. The evolution proceeds along a
family of ingoing null hypersurfaces with outer boundary at $r=60\,M$. The
evolution is stopped before the pulse hits the outer boundary in order to avoid
spurious effects from reflection and the radiation is inferred from data at
$r=20\,M$. Although gauge ambiguities arise in reading off the waveform at a
finite  radius, the work reveals interesting nonlinear effects: (i)
modification of the light cone structure governing the principal part of the
equations and hence the propagation of signals; (ii) modulation of the
Schwarzschild potential by the introduction of an angular dependent ``mass
aspect''; and (iii) quadratic and higher order terms in the evolution equations
which couple the spherical harmonic modes. A compactified version of this
study~\cite{zlochmode} was later carried out with the 3D PITT code, which
confirms these effects as well as new effects which are not present in the
axisymmetric case (see Section~\ref{sec:mode} for details).


\subsubsection{Twisting axisymmetry}
\label{sec:axiev}

The Southampton group, as part of its goal of combining Cauchy and
characteristic evolution, has developed a code~\cite{south1,south2,pollney}
which extends the Bondi problem to full axisymmetry, as described by the
general characteristic formalism of Sachs~\cite{sachs}. By dropping the
requirement that the rotational Killing vector be twist-free, they were able to
include rotational effects, including radiation in the ``cross'' polarization
mode (only the ``plus'' mode is allowed by twist-free axisymmetry). The null
equations and variables were recast into a suitably regularized form to allow
compactification of null infinity. Regularization at the vertices or caustics
of the null hypersurfaces was not necessary, since they anticipated matching to
an interior Cauchy evolution across a finite worldtube.

The code was designed to insure standard Bondi coordinate conditions at
infinity, so that the metric has the asymptotically Minkowskian form
corresponding to null-spherical coordinates. In order to achieve this, the
hypersurface equation for the Bondi metric variable $\beta$ must be integrated
radially inward from infinity, where the integration constant is specified. The
evolution of the dynamical variables proceeds radially outward as dictated by
causality~\cite{pollney}. This differs from the Pittsburgh code in which all
the equations are integrated radially outward, so that the coordinate
conditions are determined at the inner boundary and the metric is
asymptotically flat but not asymptotically Minkowskian. The Southampton scheme
simplifies the formulae for the Bondi news function and mass in terms of the
metric. It is anticipated that the inward integration of $\beta$ causes no
numerical problems because this is a gauge choice which does not propagate
physical information. However, the code has not yet been subject to convergence
and long term stability tests so that these issues cannot be properly assessed
at the present time.

The matching of the Southampton axisymmetric code
to a Cauchy interior is discussed in Section~\ref{sec:aximatch}.


\subsection{The Bondi mass}

Numerical calculations of asymptotic quantities such as the Bondi mass  must
pick off non-leading terms in an asymptotic expansion about infinity. This is
similar to the experimental task of determining the mass of an object by
measuring its far field. For example, in an asymptotically inertial frame
(called a standard Bondi frame at $\mathcal{I}^+$), the mass aspect ${\cal
M}(u,\theta,\phi)$ is picked off from the asymptotic expansion of Bondi's
metric quantity $V$ (see Equation~(\ref{eq:V})) of the form $V = r- 2\mathcal{M}
+\mathcal{O} (1/r)$. In gauges which incorporate some of the properties of an
asymptotically inertial frame, such as the null quasi-spherical
gauge~\cite{bartnumeth} in which the angular metric is conformal to the unit
sphere metric, this can be a straightforward computational problem. However,
the job can be more difficult if the gauge does not correspond to a standard
Bondi frame at $\mathcal{I}^+$. One must then deal with an arbitrary
coordinatization of $\mathcal{I}^+$ which is determined by the details of the
interior geometry. As a result, $V$ has a more complicated asymptotic behavior,
given in the axisymmetric case by
\begin{eqnarray}
  V - r &=& \frac{r^2 (L \sin \theta)_{,\theta}}{\sin \theta} +
  r e^{2 (H - K)} \times
  \nonumber \\
  && \left[ \left( 1 - e^{- 2 (H - K)} \right) +
  \frac{2 (H_{,\theta} \sin \theta)_{,\theta}}{\sin \theta} +
  K_{,\theta \theta} + 3 K_{,\theta} \cot \theta + 4 (H_{,\theta})^2 -
  4 H_{,\theta} K_{,\theta} - 2 (K_{,\theta})^2 \right]
  \nonumber \\
  && - 2 e^{2 H} \mathcal{M} + \mathcal{O} (r^{-1}),
  \label{eq:wasym}
\end{eqnarray}%
where $L$, $H$, and $K$ are gauge dependent functions of $(u,\theta)$
which would vanish in a Bondi frame~\cite{tam,isaac}. The calculation
of the Bondi mass requires regularization of this expression by
numerical techniques so that the coefficient $\mathcal{M}$ can be picked
off. The task is now similar to the experimental determination of the
mass of an object by using non-inertial instruments in a far zone which
contains $\mathcal{O} (1/r)$ radiation fields. But it has been done!

It was accomplished in Stewart's code by re-expressing the formula for
the Bondi mass in terms of the well-behaved fields of the conformal
formalism~\cite{stewbm}. In the Pittsburgh code, it was accomplished by
re-expressing the Bondi mass in terms of renormalized metric variables
which regularize all calculations at $\mathcal{I}^+$ and make them second
order accurate in grid size~\cite{mbondi}. The calculation of the Bondi news
function (which provides the waveforms of both polarization modes) is
an easier numerical task than the Bondi mass. It has also been
implemented in both of these codes, thus allowing the important check
of the Bondi mass loss formula.

An alternative approach to computing the Bondi mass is to adopt a gauge which
corresponds more closely to an inertial or Bondi frame at $\mathcal{I}^+$ and
simplifies the asymptotic limit. Such a choice is the null quasi-spherical
gauge in which the angular part of the metric is proportional to the unit
sphere metric, and as a result the gauge term $K$ vanishes in
Equation~(\ref{eq:wasym}). This gauge was adopted by Bartnik and
Norton at Canberra
in their development of a 3D characteristic evolution code~\cite{bartnumeth}
(see Section~\ref{sec:3d} for further discussion). It allowed accurate computation
of the Bondi mass as a limit as $r \rightarrow\infty$ of the Hawking
mass~\cite{bartint}.

Mainstream astrophysics is couched in Newtonian concepts, some of which
have no well defined extension to general relativity. In order to
provide a sound basis for relativistic astrophysics, it is crucial to
develop general relativistic concepts which have well defined and
useful Newtonian limits. Mass and radiation flux are
fundamental in this regard. The results of characteristic codes show
that the energy of a radiating system can be evaluated rigorously and
accurately according to the rules for asymptotically flat spacetimes,
while avoiding the deficiencies that plagued the ``pre-numerical'' era
of relativity:  (i) the use of coordinate dependent concepts such as
gravitational energy-momentum pseudotensors; (ii) a rather loose notion
of asymptotic flatness, particularly for radiative spacetimes; (iii)
the appearance of divergent integrals; and (iv) the use of
approximation formalisms, such as weak field or slow motion
expansions, whose errors have not been rigorously estimated.

Characteristic codes have extended the role of the Bondi mass from that of a
geometrical construct in the theory of isolated systems to that
of a highly accurate computational tool. The Bondi mass loss formula
provides an important global check on the preservation of the Bianchi
identities. The mass loss rates themselves have important astrophysical
significance. The numerical results demonstrate that computational
approaches, rigorously based upon the geometrical definition of mass in
general relativity, can be used to calculate radiation losses in highly
nonlinear processes where perturbation calculations would not be
meaningful.

Numerical calculation of the Bondi mass has been used to explore both
the Newtonian and the strong field limits of general
relativity~\cite{mbondi}. For a quasi-Newtonian system of radiating
dust, the numerical calculation joins smoothly on to a post-Newtonian
expansion of the energy in powers of $1/c$, beginning with the
Newtonian mass and mechanical energy as the leading terms. This
comparison with perturbation theory has been carried out to $\mathcal{O} (1/c^7)$,
at which stage the computed Bondi mass peels away from the
post-Newtonian expansion. It remains strictly positive, in contrast to
the truncated post-Newtonian behavior which leads to negative values.

A subtle feature of the Bondi mass stems from its role as one component of the
total energy-momentum 4-vector, whose calculation requires identification of
the translation subgroup of the Bondi--Metzner--Sachs group~\cite{bms}. This
introduces boost freedom into the problem. Identifying the translation subgroup
is tantamount to knowing the conformal transformation to a standard Bondi
frame~\cite{tam} in which the time slices of $\mathcal{I}$ have unit sphere
geometry. Both Stewart's code and the Pittsburgh code adapt the coordinates to
simplify the description of the interior sources. This results in a non-standard
foliation of $\mathcal{I}$. The determination of the conformal factor which
relates the 2-metric $h_{AB}$ of a slice of $\mathcal{I}$ to the unit sphere
metric is an elliptic problem equivalent to solving the second order partial
differential equation for the conformal transformation of Gaussian curvature.
In the axisymmetric case, the PDE reduces to an ODE with respect to the angle
$\theta$, which is straightforward to solve~\cite{mbondi}. The integration
constants determine the boost freedom along the axis of symmetry.

The non-axisymmetric case is more complicated. Stewart~\cite{stewbm}
proposes an approach based upon the dyad decomposition
\begin{equation}
  h_{AB} \, dx^A \, dx^B = m_A \, dx^A \, {\bar m}_B \, dx^B.
\end{equation}
The desired conformal transformation is obtained by first relating
$h_{AB}$ conformally to the flat metric of the complex plane. Denoting
the complex coordinate of the plane by $\zeta$, this relationship can
be expressed as $d\zeta = e^f m_A \, dx^A$. The conformal factor $f$ can
then be determined from the integrability condition
\begin{equation}
  m_{[A} \partial_{B]} f = \partial_{\,[A}m_{B]}.
\end{equation}
This is equivalent to the classic Beltrami equation for finding
isothermal coordinates. It would appear to be a more effective scheme
than tackling the second order PDE directly, but numerical
implementation has not yet been carried out.


\subsection{3D characteristic evolution}
\label{sec:3d}

There has been rapid progress in 3D characteristic evolution. There are now two
independent 3D codes, one developed at Canberra and the other at Pittsburgh
(the PITT code), which have the capability to study gravitational waves in
single black hole spacetimes, at a level still not mastered by Cauchy codes.
Several years ago the Pittsburgh group established robust stability and second
order accuracy of a fully nonlinear 3D code which calculates waveforms at null
infinity~\cite{cce,high} and also tracks a dynamical black hole and excises its
internal singularity from the computational grid~\cite{excise,wobb}. The
Canberra group has implemented an independent nonlinear 3D code which
accurately evolves the exterior region of a Schwarzschild black hole. Both
codes pose data on an initial null hypersurface and on a worldtube boundary,
and evolve the exterior spacetime out to a compactified version of null
infinity, where the waveform is computed. However, there are essential
differences in the underlying geometrical formalisms and numerical techniques
used in the two codes and in their success in evolving generic black hole
spacetimes.


\subsubsection{Geometrical formalism}

The PITT code uses a standard Bondi--Sachs null coordinate system,
\begin{equation}
  ds^2 =
  - \left( e^{2 \beta} \frac{V}{r} - r^2 h_{AB} \, U^A \, U^B \right) du^2 -
  2 e^{2 \beta} du \, dr - 2 r^2 h_{AB} \, U^B \, du \, dx^A +
  r^2 h_{AB} \, dx^A\, dx^B,
  \label{eq:umet}
\end{equation}
where $\det(h_{AB})=\det(q_{AB})$ for some standard choice $q_{AB}$ of the unit
sphere metric. This generalizes Equation~(\ref{eq:bmetric}) to the three-dimensional case.
The hypersurface equations derive from the ${G_\mu}^\nu \nabla_\nu u$ components
of the Einstein tensor. They take the explicit form
\begin{eqnarray}
  \beta_{,r} &=& \frac{1}{16} r \, h^{AC} \, h^{BD} \, h_{AB,r} \, h_{CD,r},
  \label{eq:3beta} \\
  \left( r^4 e^{-2 \beta} h_{AB} U^B_{,r} \right)_{,r} &=&
  2 r^4 \left( r^{-2} \beta_{,A} \right)_{,r} -
  r^2 h^{BC} D_C (h_{AB,r})
  \label{eq:3u} \\
  2 e^{-2 \beta} V_{,r} &=& \mathcal{R} - 2 D^{A} D_{A} \beta -
  2 (D^A \beta) D_A \beta + r^{-2} e^{-2 \beta} D_A
  \left( \! \left( r^4 U^A \right)_{,r} \! \right)
  \nonumber \\
  && - \frac{1}{2} r^4 e^{- 4 \beta} h_{AB} \, U^A_{,r} \, U^B_{,r},
  \label{eq:3v}
\end{eqnarray}%
where $D_A$ is the covariant derivative and $\mathcal{R}$ the curvature scalar of
the conformal 2-metric $h_{AB}$ of the $r=\mathrm{const.}$ surfaces, and capital Latin
indices are raised and lowered with $h_{AB}$. Given the null data $h_{AB}$ on
an outgoing null hypersurface, this hierarchy of equations can be integrated
radially in order to determine $\beta$, $U^A$ and $V$ on the hypersurface in
terms of integration constants on an inner boundary. The evolution equations
for the $u$-derivative of the null data derive from the trace-free part of the
angular components of the Einstein tensor, i.e.\ the components $m^A m^B G_{AB}$
where $h^{AB}=2m^{(A}\bar m^{B)}$.
They take the explicit form
\begin{eqnarray}
  m^A m^B \biggl( \!\!\!\! && ( r h_{AB,u})_{,r} -
  \frac{1}{2r} (r V h_{AB,r})_{,r} -
  \frac{2}{r} e^\beta D_A D_B e^\beta + r h_{AC} D_B (U^C_{,r})
  \nonumber \\
  && - \frac{r^3}{2} e^{-2 \beta} h_{AC} \, h_{BD} \, U^C_{,r} \, U^D_{,r} +
  2 D_A \, U_B + \frac{r}{2} h_{AB,r} \, D_C \, U^C
  \nonumber \\
  && + r U^C D_C (h_{AB,r}) + r h_{AD,r} \, h^{CD} (D_B U_C - D_C U_B)
  \biggr) = 0.
  \label{eq:hev}
\end{eqnarray}%

The Canberra code employs a null quasi-spherical (NQS) gauge (not to be
confused with the quasi-spherical approximation in which quadratically
aspherical terms are ignored~\cite{cce}). The NQS gauge takes advantage of the
possibility of mapping the angular part of the Bondi metric conformally onto a
unit sphere metric, so that $h_{AB}\rightarrow q_{AB}$. The required
transformation $x^A \rightarrow y^A(u,r,x^A)$ is in general dependent upon $u$
and $r$ so that the NQS angular coordinates $y^A$ are not constant
along the outgoing null rays, unlike the Bondi--Sachs angular coordinates.
Instead the coordinates $y^A$ display the analogue of a shift on the null
hypersurfaces $ u = \mathrm{const} $. In addition, the NQS spheres $ (u,r) =
\mathrm{const.}$ are not the same as the Bondi spheres. The radiation content of
the metric is contained in a shear vector describing this shift. This results
in the description of the radiation in terms of a spin-weight 1 field, rather
than the spin-weight 2 field associated with $h_{AB}$ in the Bondi--Sachs
formalism. In both the Bondi--Sachs and NQS gauges, the independent
gravitational data on a null hypersurface is the conformal part of its
degenerate 3-metric. The Bondi--Sachs null data consists of $h_{AB}$, which
determines the intrinsic conformal metric of the null hypersurface. In the
NQS case, $h_{AB}=q_{AB}$ and the shear vector comprises the only
non-trivial part of the conformal 3-metric. Both the Bondi--Sachs and NQS gauges
can be arranged to coincide in the special case of shear-free Robinson--Trautman
metrics~\cite{derry,bartgauge}.

The formulation of Einstein's equations in the NQS gauge is
presented in~\cite{bartee}, and the associated gauge freedom arising from
$(u,r)$ dependent rotation and boosts of the unit sphere is discussed
in~\cite{bartgauge}. As in the PITT code, the main equations involve
integrating a hierarchy of hypersurface equations along the radial null
geodesics extending from the inner boundary to null infinity. In the
NQS gauge the source terms for these radial ODE's are rather
simple when the unknowns are chosen to be the connection coefficients.
However, as a price to pay for this simplicity, after the radial integrations
are performed on each null hypersurface a first order elliptic equation must
be solved on each $ r = \mathrm{const.} $ cross-section to reconstruct
the underlying metric.


\subsubsection{Numerical methodology}

The PITT code is an explicit second order finite difference evolution algorithm
based upon retarded time steps on a uniform three-dimensional null coordinate grid.
The straightforward numerical approach and the second order convergence of the
finite difference equations has facilitated code development. The Canberra code
uses an assortment of novel and elegant numerical methods. Most of these
involve smoothing or filtering and have obvious advantage for removing short
wavelength noise but would be unsuitable for modeling shocks.

Both codes require the ability to handle tensor fields and their derivatives on
the sphere. Spherical coordinates and spherical harmonics are natural analytic
tools for the description of radiation, but their implementation in
computational work requires dealing with the impossibility of smoothly covering
the sphere with a single coordinate grid. Polar coordinate singularities in
axisymmetric systems can be regularized by standard tricks. In the absence of
symmetry, these techniques do not generalize and would be especially
prohibitive to develop for tensor fields.

A crucial ingredient of the PITT code is the {\it eth}-module~\cite{competh}
which incorporates a computational version of the Newman--Penrose
eth-formalism~\cite{eth}. The eth-module covers the sphere with two overlapping
stereographic coordinate grids (North and South). It provides everywhere
regular, second order accurate, finite difference expressions for tensor fields
on the sphere and their covariant derivatives. The eth-calculus
simplifies the underlying equations, avoids spurious coordinate singularities
and allows accurate differentiation of tensor fields on the sphere in a
computationally efficient and clean way. Its main weakness is the numerical
noise introduced by interpolations (fourth order accurate) between the North and
South patches. For parabolic or elliptic equations on the sphere, the finite
difference approach of the eth-calculus would be less efficient than a spectral
approach, but no parabolic or elliptic equations appear in the Bondi--Sachs
evolution scheme.

The Canberra code handles fields on the sphere by means of a 3-fold
representation: (i) as discretized functions on a
spherical grid uniformly spaced in standard $(\theta,\phi)$
coordinates, (ii) as fast-Fourier transforms with respect to
$(\theta,\phi)$ (based upon a smooth map of the torus onto the
sphere), and (iii) as a spectral decomposition of scalar, vector, and
tensor fields in terms of spin-weighted spherical harmonics. The grid
values are used in carrying out nonlinear algebraic operations; the
Fourier representation is used to calculate
$(\theta,\phi)$-derivatives; and the spherical harmonic representation
is used to solve global problems, such as the solution of the first
order elliptic equation for the reconstruction of the metric, whose
unique solution requires pinning down the $\ell=1$ gauge freedom. The
sizes of the grid and of the Fourier and spherical harmonic
representations are coordinated. In practice, the spherical harmonic
expansion is  carried out to 15th order in $\ell$, but the resulting
coefficients must then be projected into the $\ell \le 10$ subspace in
order to avoid inconsistencies between the spherical harmonic,
grid, and Fourier representations.

The Canberra code solves the null hypersurface equations by combining an 8th
order Runge--Kutta integration with a convolution spline to interpolate field
values. The radial grid points are dynamically positioned to approximate
ingoing null geodesics, a technique originally due to Goldwirth and
Piran~\cite{goldw} to avoid the problems with a uniform $r$-grid near a horizon
which arise from the degeneracy of an areal coordinate on a stationary horizon.
The time evolution uses the method of lines with a fourth order Runge--Kutta
integrator, which introduces further high frequency filtering.


\subsubsection{Stability}
\label{sec:stability}

\begin{description}
\item[PITT code]~\\
  Analytic stability analysis  of the finite difference equations has
  been crucial in the development of a stable evolution algorithm,
  subject to the standard Courant--Friedrichs--Lewy (CFL) condition
  for an explicit  code. Linear stability analysis on Minkowski and
  Schwarzschild backgrounds showed that certain field variables must
  be represented on the half-grid~\cite{papa,cce}. Nonlinear stability
  analysis was essential in revealing and curing a mode coupling
  instability that was not present in the original axisymmetric
  version of the code~\cite{high,luisdis}. This has led to a code
  whose stability persists even in the regime that the $u$-direction,
  along which the grid flows, becomes spacelike, such as outside the
  velocity of light cone in a rotating coordinate system. Severe tests
  were used to verify stability. In the linear regime, \emph{robust
    stability} was established by imposing random initial data on the
  initial characteristic hypersurface and random constraint violating
  boundary data on an inner worldtube. (Robust stability was later
  adopted as one of the standardized tests for Cauchy
  codes~\cite{apples}.) The code ran stably for 10,000 grid crossing
  times under these conditions~\cite{cce}. The PITT code was the first
  3D general relativistic code to pass this robust stability test. The
  use of random data is only possible in sufficiently weak cases where
  effective energy terms quadratic in the field gradients are not
  dominant. Stability in the highly nonlinear regime was tested in two
  ways. Runs for a time of $ 60,000 \, M$ were carried out for a moving,
  distorted Schwarzschild black hole (of mass $M$), with the
  marginally trapped surface at the inner boundary tracked and its
  interior excised from the computational
  grid~\cite{wobb,stablett}. This remains one of the longest
  simulations of a dynamic black hole carried out to
  date. Furthermore, the scattering of a gravitational wave off a
  Schwarzschild black hole was successfully carried out in the extreme
  nonlinear regime where the backscattered Bondi news was as large as
  $N=400$ (in dimensionless geometric units)~\cite{high}, showing that
  the code can cope with the enormous power output $N^2 c^5/G \approx
  10^{60} \mathrm{\ W}$ in conventional units. This exceeds the power
  that would be produced if, in 1 second, the entire galaxy were
  converted to gravitational radiation.
\end{description}

\begin{description}
\item[Canberra code]~\\
  Analytic stability analysis of the underlying finite difference
  equations is impractical because of the extensive mix of spectral
  techniques, higher order methods, and splines. Although there is no
  clear-cut CFL limit on the code, stability tests show that there is
  a limit on the time step. The damping of high frequency modes due to
  the implicit filtering would be expected to suppress numerical
  instability, but the stability of the Canberra code is nevertheless
  subject to two qualifications~\cite{bartacc,bartnumsol,bartnumeth}:
  (i) At late times (less than $100 \, M$), the evolution terminates as
  it approaches an event horizon, apparently because of a breakdown of
  the NQS gauge condition, although an analysis of how and why this
  should occur has not yet been given. (ii) Numerical instabilities
  arise from dynamic inner boundary conditions and restrict the inner
  boundary to a fixed  Schwarzschild horizon. Tests in the extreme
  nonlinear regime were not reported.
\end{description}


\subsubsection{Accuracy}

\begin{description}
\item[PITT code]~\\
  Second order accuracy has been verified in an extensive number of
  testbeds~\cite{cce,high,wobb,Zlochower,zlochmode}, including new
  exact solutions specifically constructed in null coordinates for the
  purpose of convergence tests:
  \begin{itemize}
  \item Linearized waves on a Minkowski background in null cone
    coordinates.
  \item Boost and rotation symmetric solutions~\cite{boostrot}.
  \item Schwarzschild in rotating coordinates.
  \item Polarization symmetry of nonlinear twist-free axisymmetric
    waveforms.
  \item Robinson--Trautman waveforms from perturbed Schwarzschild black
    holes.
  \item Nonlinear Robinson--Trautman waveforms utilizing an
    independently computed solution of the Robinson--Trautman equation.
  \item Perturbations of a Schwarzschild black hole utilizing an
    independent computed solution of the Teukolsky equation.
  \end{itemize}
\end{description}

\begin{description}
\item[Canberra code]~\\
  The complexity of the algorithm and NQS gauge makes it problematic
  to establish accuracy by direct means. Exact solutions do not
  provide an effective convergence check, because the Schwarzschild
  solution is trivial in the NQS gauge and other known solutions in
  this gauge require dynamic inner boundary conditions which
  destabilize the present version of the code. Convergence to
  linearized solutions is a possible check but has not yet been
  performed. Instead indirect tests by means of geometric consistency
  and partial convergence tests are used to calibrate accuracy. The
  consistency tests were based on the constraint equations, which are
  not enforced during null evolution except at the inner boundary. The
  balance between mass loss and radiation flux through $\mathcal{I}^+$
  is a global consequence of these constraints. No appreciable growth
  of the constraints was noticeable until within $5 \, M$ of the final
  breakdown of the code. In weak field tests where angular resolution
  does not dominate the error, partial convergence tests based upon
  varying the radial grid size verify the 8th order convergence in the
  shear expected from the Runge--Kutta integration and splines. When
  the radial source of error is small, reduced error with smaller time
  step can also be discerned.
  
  In practical runs, the major source of inaccuracy is the spherical
  harmonic resolution, which was restricted to $\ell \le 15$ by
  hardware limitations. Truncation of the spherical harmonic expansion
  has the effect of modifying the equations to a system for which the
  constraints are no longer satisfied. The relative error in the
  constraints is $10^{-3}$ for strong field
  simulations~\cite{bartint}.
\end{description}


\subsubsection{First versus second differential order}

The PITT code was originally formulated in the second differential
form of Equations~(\ref{eq:3beta}, \ref{eq:3u}, \ref{eq:3v},
\ref{eq:hev}), which in the spin-weighted version leads to
an economical number of 2 real and 2 complex variables. Subsequently, the
variable
\begin{equation}
  Q_A = r^2 e^{-2 \beta} h_{AB} \, U^B_{,r}
\end{equation}
was introduced to reduce Equation~(\ref{eq:3u}) to two first order radial equations,
which simplified the startup procedure at the initial boundary. Although the
resulting code has been verified to be stable and second order accurate, its
application to increasingly difficult problems involving strong fields, and
gradients have led to numerical errors that make important physical effects
hard to measure. In particular, in initial attempts to simulate a white hole
fission, G{\'{o}}mez~\cite{gomezfo} encountered an oscillatory error pattern in
the angular directions near the time of fission. The origin of the
problem was
tracked to numerical error of an oscillatory nature introduced by $\eth^2$
terms in the hypersurface and evolution equations. G{\'{o}}mez' solution was
to remove the offending second angular derivatives by introducing additional
variables and reducing the system to first differential order in the angular
directions. This suppressed the oscillatory mode and subsequently improved
performance in the simulation of the white hole fission problem~\cite{fiss}
(see Section~\ref{sec:fission}).

This success opens the issue of whether a completely first differential order
code might perform even better, as has been proposed by G{\'{o}}mez and
Frittelli~\cite{gomfrit}. They gave a first order quasi-linear formulation of
the Bondi system which, at the analytic level, obeys a standard uniqueness and
existence theorem (extending previous work for the linearized
case~\cite{lehnfrit}); and they point out, at the numerical level, that a
first order code could also benefit from the applicability of standard
numerical techniques. This is an important issue which is not simple to resolve
without code comparison. The part of the code in which the $\eth^2$ operator
introduced the  oscillatory error mode was not identified in~\cite{gomezfo},
i.e.\ whether it originated in the inner boundary treatment or in the
interpolations between stereographic patches where second derivatives might be
troublesome. There are other possible ways to remove the oscillatory angular
modes, such as adding angular dissipation or using more accurate methods of
patching the sphere. The current finite difference algorithm only introduces
numerical dissipation in the radial direction~\cite{luisdis}. The economy of
variables in the original Bondi scheme should not be abandoned without further
tests and investigation.


\subsubsection{Nonlinear scattering off a Schwarzschild black hole}

A natural physical application of a characteristic evolution code is the
nonlinear version of the classic problem of scattering off a Schwarzschild
black hole, first solved perturbatively by Price~\cite{price}. Here the inner
worldtube for the characteristic initial value problem consists of the ingoing
branch of the $r=2m$ hypersurface (the past horizon), where Schwarzschild data
are prescribed. The nonlinear problem of a gravitational wave scattering off a
Schwarzschild black hole is then posed in terms of data on an outgoing null
cone which describe an incoming pulse with compact support. Part of the energy
of this pulse falls into the black hole and part is backscattered to ${\cal
I}^+$. This problem has been investigated using both the PITT and Canberra
codes.

The Pittsburgh group studied the backscattered waveform (described by the Bondi
news function) as a function of incoming pulse amplitude. The
computational eth-module smoothly handled the complicated time
dependent transformation between
the non-inertial computational frame at $\mathcal{I}^+$ and the inertial (Bondi)
frame necessary to obtain the standard ``plus'' and ``cross'' polarization
modes. In the perturbative regime, the news corresponds to the backscattering
of the incoming pulse off the effective Schwarzschild potential. When the
energy of the pulse is no larger than the central Schwarzschild mass,  the
backscattered waveform  still depends roughly linearly on the
amplitude of the incoming pulse. However, for very high amplitudes the
waveform behaves quite differently. Its amplitude is greater than that
predicted by linear scaling and its shape drastically changes and exhibits
extra oscillations. In this very high amplitude case, the mass of the system is
completely dominated by the incoming pulse, which essentially backscatters off
itself in a nonlinear way.

The Canberra code was used to study the change in Bondi mass due to the
radiation~\cite{bartint}. The Hawking mass $M_\mathrm{H}(u,r)$ was calculated as a
function of radius and retarded time, with the Bondi mass $M_\mathrm{B}(u)$ then
obtained by taking the limit $r\rightarrow \infty$. The limit had good
numerical behavior. For a strong initial pulse with $\ell=4$ angular
dependence, in a run from $u=0$ to $u=70$ (in units where the interior
Schwarzschild mass is 1), the Bondi mass dropped from 1.8 to 1.00002, showing
that almost half of the initial energy of the system was backscattered and that
a surprisingly negligible amount of energy fell into the black hole. A possible
explanation is that the truncation of the spherical harmonic expansion cuts off
wavelengths small enough to effectively penetrate the horizon. The Bondi mass
decreased monotonically in time, as necessary theoretically, but its rate of
change exhibited an interesting pulsing behavior whose time scale could not be
obviously explained in terms of quasinormal oscillations. The Bondi mass loss
formula was confirmed with relative error of less than $10^{-3} $. This is
impressive accuracy considering the potential sources of numerical error
introduced by taking the limit of the Hawking mass with limited resolution. The
code was also used to study the appearance of logarithmic terms in the
asymptotic expansion of the Weyl tensor~\cite{bartnort}. In addition, the
Canberra group studied the effect of the initial pulse amplitude on the waveform of
the backscattered radiation, but did not extend their study to the very high
amplitude regime in which qualitatively interesting nonlinear effects occur.


\subsubsection{Black hole in a box}

The PITT code has also been implemented to evolve along an advanced time
foliation by \emph{ingoing} null cones, with data given on a worldtube at their
\emph{outer} boundary and on the initial \emph{ingoing} null cone. The code was
used to evolve a black hole in the region interior to the worldtube by
implementing a horizon finder to locate the marginally trapped surface (MTS) on
the ingoing cones and excising its singular interior~\cite{excise}. The code
tracks the motion of the MTS and measures its area during the evolution. It was
used to simulate a distorted ``black hole in a box''~\cite{wobb}. Data at the
outer worldtube was induced from a Schwarzschild or Kerr spacetime but the
worldtube was allowed to move relative to the stationary trajectories; i.e.\
with respect to the grid the worldtube is fixed but the black hole moves inside
it. The initial null data consisted of a pulse of radiation which subsequently
travels outward to the worldtube where it reflects back toward the black hole.
The approach of the system to equilibrium was monitored by the area of the MTS,
which also equals its Hawking mass. When the worldtube is stationary (static or
rotating in place), the distorted black hole inside evolved to equilibrium with
the boundary. A boost or other motion of the worldtube with respect to the
black hole did not affect this result. The marginally trapped surface always
reached equilibrium with the outer boundary, confirming that the motion of the
boundary was ``pure gauge''.

The code runs ``forever'' even when the worldtube wobbles with respect to the
black hole to produce artificial periodic time dependence. An initially
distorted, wobbling black hole was evolved for a time of $60,000 \, M$, longer by
orders of magnitude than permitted by the stability of other existing 3D black
hole codes at the time. This exceptional performance opens a promising new
approach to handle the inner boundary condition for Cauchy evolution of black
holes by the matching methods reviewed in Section~\ref{sec:ccm}.

Note that setting the pulse to zero is equivalent to prescribing shear free
data on the initial null cone. Combined with Schwarzschild boundary data on the
outer world tube, this would be complete data for a Schwarzschild space time.
However, the evolution of such shear free null data combined with Kerr boundary
data would have an initial transient phase before settling down to a Kerr black
hole. This is because the twist of the shear-free Kerr null congruence implies
that Kerr data specified on a null hypersurface are not generally shear free.
The event horizon is an exception but Kerr null data on other null
hypersurfaces have not been cast in explicit analytic form. This makes the Kerr
spacetime an awkward testbed for characteristic codes. (Curiously, Kerr data on
a null hypersurface with a conical type singularity do take a simple analytic
form, although unsuitable for numerical evolution~\cite{lun}.) Using some
intermediate analytic results of Israel and Pretorius~\cite{IsrPret}, Venter
and Bishop~\cite{bishkerr} have recently constructed a numerical algorithm for
transforming the Kerr solution into Bondi coordinates and in that way provide
the necessary null data numerically.


\subsection{Characteristic treatment of binary black holes}
\label{sec:nullbbh}

An important application of characteristic evolution is the calculation of the
waveform emitted by binary black holes, which is possible during the very
interesting nonlinear domain from merger to ringdown~\cite{ndata,kyoto}. The
evolution is carried out along a family of ingoing null hypersurfaces which
intersect the horizon in topological spheres. It is restricted to the period
following the merger, for otherwise the ingoing null hypersurfaces would
intersect the horizon in disjoint pieces corresponding to the individual black
holes. The evolution proceeds \emph{backward} in time on an ingoing null
foliation to determine the exterior spacetime in the post-merger era. It is an
example of the characteristic initial value problem posed on an intersecting
pair of null hypersurfaces~\cite{sachsdn,haywdn}, for which existence theorems
apply in some neighborhood of the initial null
hypersurfaces~\cite{hagenseifert77,helmut81a,fried}. Here one of the null
hypersurfaces is the event horizon $\mathcal{H}^+$ of the binary black holes. The
other is an ingoing null hypersurface $J^+$ which intersects $\mathcal{H}^+$ in a
topologically spherical surface $\mathcal{S}^+$ approximating the equilibrium of
the final Kerr black hole, so that $J^+$ approximates future null infinity
$\mathcal{I}^+$. The required data for the analytic problem consists of the
degenerate conformal null metrics of $\mathcal{H}^+$ and $J^+$ and the metric and
extrinsic curvature of their intersection $\mathcal{S}^+$.

The conformal metric of $\mathcal{H}^+$ is provided by the conformal horizon model
for a binary black hole horizon~\cite{ndata,asym}, which treats the horizon in
stand-alone fashion as a three-dimensional manifold endowed with a degenerate
metric $\gamma_{ab}$ and affine parameter $t$ along its null rays. The metric
is obtained from the conformal mapping $\gamma_{ab}=\Omega^2 \hat \gamma_{ab}$
of the intrinsic metric $\hat \gamma_{ab}$ of a flat space null hypersurface
emanating from a convex surface $\mathcal{S}_0$ embedded at constant time in
Minkowski space. The horizon is identified with the null hypersurface formed
by the inner branch of the boundary of the past of $\mathcal{S}_0$, and its
extension into the future. The flat space null hypersurface expands forever as
its affine parameter $\hat t$ (Minkowski time) increases, but the conformal
factor is chosen to stop the expansion so that the cross-sectional area of the
black hole approaches a finite limit in the future. At the same time, the
Raychaudhuri equation (which governs the growth of surface area) forces a
nonlinear relation between the affine parameters $t$ and $\hat t$. This is what
produces the nontrivial topology of the affine $t$-slices of the black hole
horizon. The relative distortion between the affine parameters $t$ and $\hat
t$, brought about by curved space focusing, gives rise to the trousers shape of
a binary black hole horizon.

\begin{figure}[hptb]
  \def\epsfsize#1#2{0.4#1}
  \centerline{\epsfbox{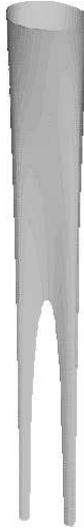}}
  \caption{\it Trousers shaped event horizon obtained by the
    conformal model.}
  \label{fig:pants}
\end{figure}

\begin{figure}[hptb]
  \def\epsfsize#1#2{0.8#1}
  \centerline{\epsfbox{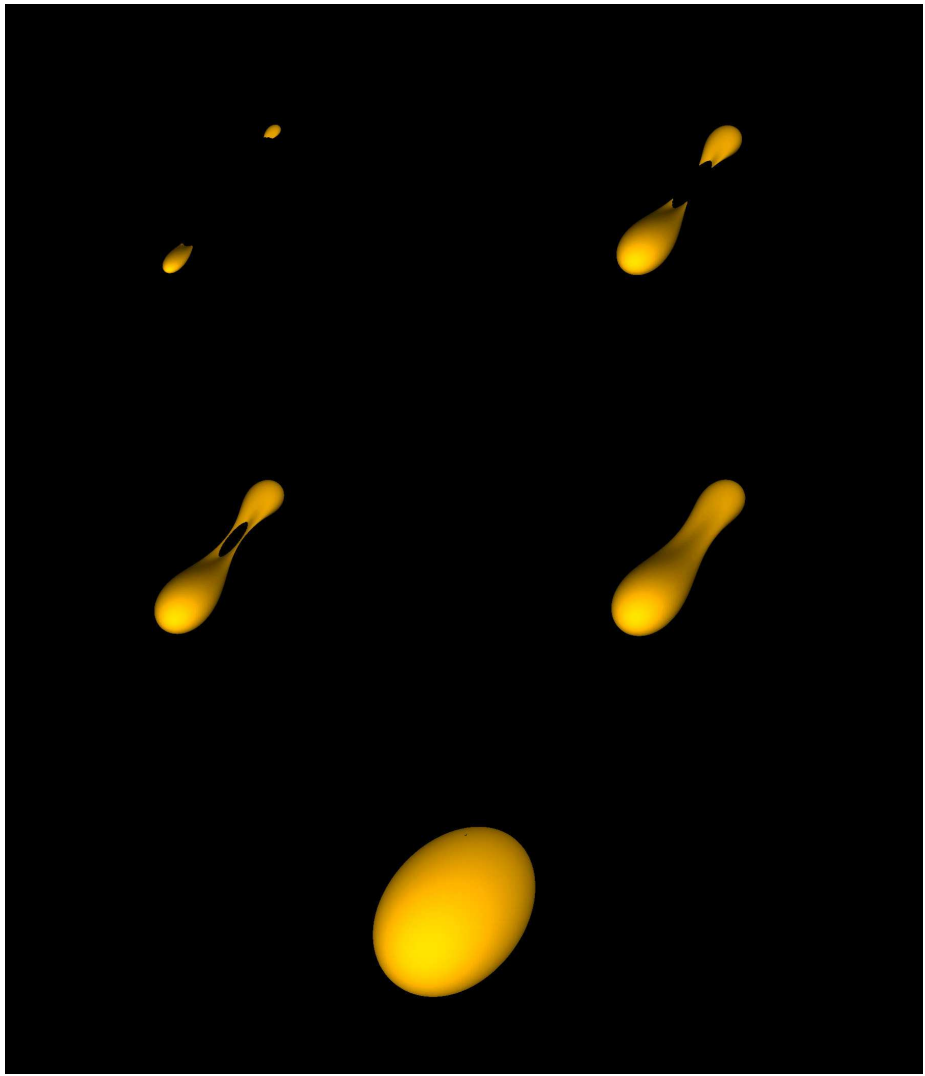}}
  \caption{\it Upper left: Tidal distortion of approaching black holes
    Upper right: Formation of sharp pincers just prior to
    merger. Middle left: Temporarily toroidal stage just after
    merger. Middle right: Peanut shaped black hole after the hole in
    the torus closes. Lower: Approach to final equilibrium.}
  \label{fig:bbh2}
\end{figure}

An embedding diagram of the horizon for an axisymmetric head-on collision,
obtained by choosing $\mathcal{S}_0$ to be a prolate spheroid, is shown in
Figure~\ref{fig:pants}~\cite{ndata}. The black hole event horizon associated with
a triaxial ellipsoid  reveals new features not seen in the  degenerate case of
the head-on collision~\cite{asym}, as depicted in Figure~\ref{fig:bbh2}. If the
degeneracy is slightly broken, the individual black holes form with spherical
topology but as they approach, tidal distortion produces two sharp pincers on
each black hole just prior to merger. At merger, the two pincers join to form a
single temporarily toroidal black hole. The inner hole of the torus
subsequently closes up to produce first a peanut shaped black hole and finally
a spherical black hole. No violation of topological censorship~\cite{fsw}
occurs because the hole in the torus closes up superluminally. Consequently, a
causal curve passing through the torus at a given time can be slipped below the
bottom of a trouser leg to yield a causal curve lying entirely outside the
hole~\cite{toroid}. In the degenerate axisymmetric limit, the pincers reduce to
a point so that the individual holes have teardrop shape and they merge without
a toroidal phase. Animations of this merger can be viewed at~\cite{psc}.

The conformal horizon model determines the data on $\mathcal{H}^+$ and ${\cal
S}^+$. The remaining data necessary to evolve the exterior spacetime are given
by the conformal geometry of $J^+$, which constitutes the outgoing radiation
waveform. The determination of the merger-ringdown waveform proceeds in two
stages. In the first stage, this outgoing waveform is set to zero and the
spacetime is evolved backward in time to calculate the incoming radiation
entering from $\mathcal{I}^-$. (This incoming radiation is eventually absorbed by
the black hole.) From a time reversed point of view, this evolution describes
the outgoing waveform emitted in the fission of a white hole, with the
physically correct initial condition of no ingoing radiation. Preliminary
calculations show that at late times the waveform is entirely quadrupolar
($\ell =2$) but that a strong octopole mode ($\ell =4$) exists just before
fission. In the second stage of the calculation, this waveform could be used to
generate the physically correct outgoing waveform for a black hole merger. The
passage from the first stage to the second is the nonlinear equivalent of first
determining an inhomogeneous solution to a linear problem and then adding the
appropriate homogeneous solution to satisfy the boundary conditions. In this
context, the first stage supplies an advanced solution and the second stage the
homogeneous retarded minus advanced solution. When the evolution is carried out
in the perturbative regime of a Kerr or Schwarzschild background, as in the
close approximation~\cite{jorge}, this superposition of solutions is simplified
by the time reflection symmetry~\cite{kyoto}. The second stage has been carried
out in the perturbative regime of the close approximation using a
characteristic code which solves the Teukolsky equation, as described in
Section~\ref{sec:schwpert}. More generally, beyond the perturbative regime, the
merger-ringdown waveform must be obtained by a more complicated inverse
scattering procedure, which has not yet been attempted.

There is a complication in applying the PITT code to this double null evolution
because a dynamic horizon does not lie precisely on $r$-grid points. As a
result, the $r$-derivative of the null data, i.e.\ the ingoing shear of ${\cal
H}$, must also be provided in order to initiate the radial hypersurface
integrations. The ingoing shear is part of the free data specified at ${\cal
S}^+$. Its value on $\mathcal{H}$ can be determined by integrating (backward in
time) a sequence of propagation equations involving the horizon's twist and
ingoing divergence. A horizon code which carries out these integrations has
been tested to give accurate data even beyond the merger~\cite{hdata}.

The code has revealed new global properties of the head-on collision by
studying a sequence of data for a family of colliding black holes  which
approaches a single Schwarzschild black hole. The resulting perturbed
Schwarzschild horizon provides global insight into the close
limit~\cite{jorge}, in which the individual black holes have joined in the
infinite past. A marginally anti-trapped surface divides the horizon into
interior and exterior regions, analogous to the division of the Schwarzschild
horizon by the $r=2M$ bifurcation sphere. In passing from the perturbative to
the strongly nonlinear regime there is a rapid transition in which the
individual black holes move into the exterior portion of the horizon. The data
paves the way for the PITT code to calculate whether this dramatic time
dependence of the horizon produces an equally dramatic waveform. See
Section~\ref{sec:fission} for first stage results.


\subsection{Perturbations of Schwarzschild}
\label{sec:schwpert}

The nonlinear 3D PITT code has been calibrated in the regime of small
perturbations of a Schwarz\-schild spacetime~\cite{Zlochower,zlochmode} by
measuring convergence with respect to independent solutions of the Teukolsky
equation~\cite{teuk72}. By decomposition into spherical harmonics,  the
Teukolsky equation reduces the problem of a perturbation of a stationary black
hole to a 1D problem in the $(t,r)$ subspace perturbations for a component of
the Weyl tensor. Historically, the Teukolsky equation was first solved
numerically by Cauchy evolution. Campanelli, G{\'{o}}mez, Husa, Winicour, and
Zlochower~\cite{zlochadv,zlochret} have reformulated the Teukolsky formalism as
a double-null characteristic evolution algorithm. The evolution proceeds on a
family of outgoing null hypersurfaces with an ingoing null hypersurface as
inner boundary and with the outer boundary compactified at future null
infinity. It applies to either the Weyl component $\Psi_0$ or $\Psi_4$, as
classified in the Newman--Penrose formalism. The $\Psi_0$ component comprises
constraint-free gravitational data on an outgoing null hypersurface and
$\Psi_4$ comprises the corresponding data  on an ingoing null hypersurface. In
the study of perturbations of a Schwarzschild black hole, $\Psi_0$ is
prescribed on an outgoing null hypersurface $\mathcal{J}^-$, representing an early
retarded time approximating past null infinity, and $\Psi_4$ is prescribed on
the inner white hole horizon $\mathcal{H}^-$.

\begin{figure}[hptb]
  \def\epsfsize#1#2{0.4#1}
  \centerline{\epsfbox{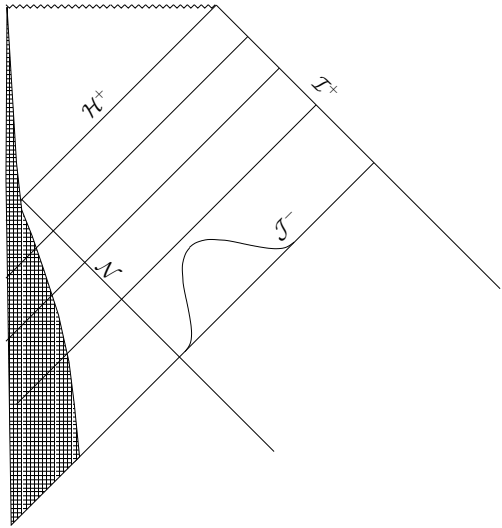}}
  \caption{\it The physical setup for the scattering problem. A star
    of mass $M$ has undergone spherically symmetric collapse to form
    a black hole. The ingoing null worldtube $\mathcal{N}$ lies outside
    the collapsing matter. Inside $\mathcal{N}$ (but outside the
    matter) there is a vacuum Schwarzschild metric. Outside of
    $\mathcal{N}$, data for an ingoing pulse is specified on the
    initial outgoing null hypersurface $\mathcal{J}^-$. As the pulse
    propagates to the black hole event horizon $\mathcal{H}^+$, part of
    its energy is scattered to $\mathcal{I}^+$.}
  \label{fig:setup}
\end{figure}

The physical setup is described in
Figure~\ref{fig:setup}. The outgoing null hypersurfaces extend to future
null infinity $\mathcal{I}^+$ on a compactified numerical grid.
Consequently, there is no need for either an artificial outer boundary
condition or an interior extraction worldtube. The outgoing radiation
is computed in the coordinates of an observer in an inertial frame at
infinity, thus avoiding any gauge ambiguity in the waveform.

The first calculations were carried out with nonzero data for $\Psi_4$  on
$\mathcal{H}^-$ and zero data on $\mathcal{J}^-$~\cite{zlochadv} (so that no ingoing
radiation entered the system). The resulting simulations were highly accurate
and tracked the quasi-normal ringdown of a perturbation consisting of a compact
pulse through 10 orders of magnitude and tracked the final power law decay
through an additional 6 orders of magnitude. The measured exponent of the power
law decay varied from $\approx 5.8$, at the beginning of the tail, to $\approx
5.9$ near the end, in good agreement with the predicted value of $2\ell+2 =6$
for a quadrupole wave~\cite{price}.

The accuracy of the perturbative solutions provide a virtual exact solution
for carrying out convergence tests of the nonlinear PITT null code. In this
way, the error in the Bondi news function computed by the PITT
code was calibrated for perturbative data consisting of either an outgoing
pulse on $\mathcal{H}^-$ or an ingoing pulse on $\mathcal{J}^-$.
For the outgoing pulse, clean second order convergence
was confirmed until late times in the evolution, when small deviations
from second order arise from accumulation of roundoff and truncation error.
For the Bondi news produced by the scattering of an ingoing
pulse, clean second order convergence was again confirmed until late times
when the pulse approached the $r=2M$ black hole horizon. The late time error
arises from loss of resolution of the pulse (in the radial direction) resulting
from the properties of the compactified radial coordinate used in the code.
This type of error could be eliminated by using characteristic AMR
techniques under development~\cite{pretlehn}.


\subsubsection{Close approximation white hole and black hole waveforms}
\label{sec:close}

The characteristic Teukolsky code has been used to study radiation from
axisymmetric white holes and black holes in the close approximation. The
radiation from an axisymmetric fissioning white hole~\cite{zlochadv} was
computed using the Weyl data on $\mathcal{H}^-$ supplied by the conformal horizon
model described in Section~\ref{sec:nullbbh}, with the fission occurring along the
axis of symmetry. The close approximation implies that the fission takes place
far in the future, i.e.\ in the region of $\mathcal{H}^-$ above the black hole
horizon $\mathcal{H}^+$. The data have a free parameter $\eta$ which controls the
energy yielded by the white hole fission. The radiation waveform reveals an
interesting dependence on the parameter $\eta$. In the large $\eta$ limit,
the  waveform consists of a single pulse, followed by ringdown and tail decay.
The amplitude of the pulse scales quadratically with $\eta$ and the width
decreases with $\eta$. As $\eta$ is reduced, the initial pulse broadens and
develops more structure. In the small $\eta$ limit, the amplitude scales
linearly with $\eta$ and the shape is independent of $\eta$.

Since there was no incoming radiation, the above model gave the physically
appropriate boundary conditions for a white hole fission (in the close
approximation). From a time reversed view point, the system corresponds to a
black hole merger with no outgoing radiation at future null infinity, i.e.\ the
analog of an advanced solution with only ingoing but no outgoing radiation. In
the axisymmetric case studied, the merger corresponds to a head-on collision
between two black holes. The physically appropriate boundary conditions for a
black hole merger correspond to no ingoing radiation on $\mathcal{J}^-$ and
binary black hole data on $\mathcal{H}^+$. Because  $\mathcal{J}^-$ and $\mathcal{H}^+$
are disjoint, the corresponding data cannot be used directly to formulate a
double null characteristic initial value problem. However, the ingoing
radiation at $\mathcal{J}^-$ supplied by the advanced solution for the black hole
merger could be used as Stage~I of a two stage approach to determine the
corresponding retarded solution. In Stage~II, this ingoing radiation is used to
generate the analogue of an \emph{advanced minus retarded} solution. A pure
retarded solution (with no ingoing radiation but outgoing radiation at ${\cal
I}^+$) can then be constructed by superposition. The time reflection symmetry
of the Schwarzschild background is key to carrying out this construction.

This two stage strategy has been carried out by Husa, Zlochower, G{\'{o}}mez, and
Winicour~\cite{zlochret}. The superposition of the Stage~I and~II solutions
removes the ingoing radiation from $\mathcal{J}^-$ while modifying the close
approximation perturbation of $\mathcal{H}^+$, essentially making it ring. The
amplitude  of the radiation waveform at $\mathcal{I}^+$ has a linear dependence on
the parameter $\eta$, which in this black hole scenario governs the energy lost
in the inelastic merger process. Unlike the fission waveforms, there is very
little $\eta$-dependence in their shape and the amplitude continues to scale
linearly even for large $\eta$. It is not surprising that the retarded
waveforms from a black hole merger differs markedly from the retarded waveforms
from a white hole merger. The white hole process is directly visible at  ${\cal
I}^+$ whereas the merger waveform results indirectly from the black holes
through the preceding collapse of matter or gravitational energy that formed
them. This explains why the fission waveform is more sensitive to the parameter
$\eta$ which controls the shape and timescale of the horizon data. However, the
weakness of the dependence of the merger waveform on $\eta$ is surprising and
has potential importance for enabling the design of an efficient template for
extracting a gravitational wave signal from noise.


\subsubsection{Fissioning white hole}
\label{sec:fission}

In the purely vacuum approach to the binary black hole problem, the stars which
collapse to form the black holes are replaced by imploding gravitational waves.
This avoids hydrodynamic difficulties at the expense of a globally complicated
initial value problem. The imploding waves either emanate from a past
singularity, in which case the time-reversed application of cosmic censorship
implies the existence of an anti-trapped surface; or they emanate from ${\cal
I}^-$, which complicates the issue of gravitational radiation content in
the initial data and its effect on the outgoing waveform. These complications
are avoided in the two stage approach adopted in the close approximation
studies described in Section~\ref{sec:close}, where advanced and retarded
solutions in a Schwarzschild background can be rigorously identified and
superimposed. Computational experiments have been carried out to study the
applicability of this approach in the nonlinear regime~\cite{fiss}.

From a time reversed viewpoint, the first stage is equivalent to the
determination of the outgoing radiation from a fission of a white hole in the
absence of ingoing radiation, i.e.\ the physically appropriate ``retarded''
waveform from a white hole fission. This fission problem can be formulated in
terms of data on the white hole horizon $\mathcal{H}^-$ and data representing
the absence of ingoing radiation on a null hypersurface $J^-$ which emanates
from $\mathcal{H}^-$ at an early time. The data on $\mathcal{H}^-$ is provided by the
conformal horizon model for a fissioning white hole. This allows study of a
range of models extending from the perturbative close approximation regime, in
which the fission occurs inside a black hole event horizon, to the
nonlinear regime of a ``bare'' fission visible from $\mathcal{I}^+$. The study
concentrates on the axisymmetric spinless fission (corresponding in the time
reversed view to the head-on collision of non-spinning black holes). In the
perturbative regime, the news function agrees with the close approximation
waveforms. In the highly nonlinear regime, a bare fission was found to produce
a dramatically sharp radiation pulse, which then undergoes a damped
oscillation. Because the fission is visible from $\mathcal{I}^+$, it is a more
efficient source of gravitational waves than a black hole merger and can
produce a higher fractional mass loss.


\subsection{Nonlinear mode coupling}
\label{sec:mode}

The PITT code has been used to  model the nonlinear generation of waveforms by
scattering off a Schwarzschild black hole~\cite{Zlochower,zlochmode}. The
physical setup is similar to the perturbative study in Section~\ref{sec:schwpert}.
A radially compact pulse is prescribed on an early time outgoing null
hypersurface $\mathcal{J}^-$ and Schwarzschild null data is given on the interior
white hole horizon $\mathcal{H}^-$, which is causally unaffected by the pulse.
The input pulse is standardized to ($\ell=2$, $m=0$) and ($\ell=2$, $m=2$)
quadrupole modes with amplitude $A$. The outgoing null hypersurfaces extend to
future null infinity $\mathcal{I}^+$ on a compactified numerical grid.
Consequently, there is no need for an artificial outer boundary. The evolution
code then provides the news function at $\mathcal{I}^+$, in the coordinates of an
observer in an inertial frame at infinity, thus avoiding any gauge ambiguity in
the waveform. This provides a simple setting how the nonlinearities generated
by high amplitudes affect the waveform.

The study reveals several features of qualitative importance:
\begin{enumerate}
\item The mode coupling amplitudes consistently scale as powers $A^n$
  of the input amplitude $A$ corresponding to the nonlinear order of
  the terms in the evolution equations which produce the mode. This
  allows much economy in producing a waveform catalog: Given the order
  $n$ associated with a given mode generation, the response to any
  input amplitude $A$ can be obtained from the response to a single
  reference amplitude.
  \label{feature_1}
\item The frequency response has similar behavior but in a less
  consistent way. The dominant frequencies produced by mode coupling
  are in the approximate range of the quasinormal frequency of the
  input mode and the expected sums and difference frequencies
  generated by the order of nonlinearity.
  \label{feature_2}
\item Large phase shifts, ranging up 15\% in a half cycle relative to
  the linearized waveform, are exhibited in the news function obtained
  by the superposition of all output modes, i.e.\ in the  waveform of
  observational significance. These phase shifts, which are important
  for design of signal extraction templates, arise in an erratic way
  from superposing modes with different oscillation frequencies. This
  furnishes a strong argument for going beyond the linearized
  approximation in designing a waveform catalog for signal extraction.
  \label{feature_3}
\item Besides the nonlinear generation of harmonic modes absent in the
  initial data, there is also a stronger than linear generation of
  gravitational wave output. This provides a potential mechanism for
  enhancing the strength of the gravitational radiation produced
  during, say, the merger phase of a binary inspiral above the
  strength predicted in linearized theory.
  \label{feature_4}
\item In the non-axisymmetric $m=2$ case, there is also considerable
  generation of radiation in polarization states not present in the
  linearized approximation. In the simulations, input amplitudes in
  the range  $A=0.1$ to $A=0.36$ lead to nonlinear generation of the
  $\oplus$ polarization mode which is of the same order of magnitude
  as the $\otimes$ mode (which would be the sole polarization in the
  linearized regime). As a result, significant nonlinear amplification
  and phase shifting of the waveform would be observed by a
  gravitational wave detector, depending on its orientation.
  \label{feature_5}
\end{enumerate}

These effects arise from the nonlinear modification of the Schwarzschild
geometry identified by Papadopoulos in his prior work on axisymmetric mode
coupling~\cite{papamode}, reported in Section~\ref{sec:papamode}. Although
Papadopoulos studied nonlinear mode generation produced by an outgoing pulse,
as opposed to the case of an ingoing pulse studied
in~\cite{Zlochower,zlochmode}, the same nonlinear factors were in play and gave
rise to several common features. In both cases, the major effects arise in the
region near $r=3M$. Analogs of Features~\ref{feature_1},
\ref{feature_2}, \ref{feature_3}, and~\ref{feature_4} above are all apparent
in Papadopoulos's work. At the finite difference level, both codes respect the
reflection symmetry inherent in Einstein's equations and exhibit the
corresponding selection rules arising from parity considerations. In the
axisymmetric case considered by Papadopoulos, this forbids the nonlinear
generation of a $\oplus$ mode  from a $\otimes$ mode, as described in
Feature~\ref{feature_5} above.

The evolution along ingoing null hypersurfaces in the axisymmetric work of
Papadopoulos has complementary numerical features with the evolution along
outgoing null hypersurfaces in the 3D work. The grid based upon \emph{ingoing}
null hypersurfaces avoids the difficulty in resolving effects close to $r=2M$
encountered with the grid based upon \emph{outgoing} null hypersurfaces. The
outgoing code would require AMR in order to resolve
the quasinormal ringdown for as many cycles as achieved by Papadopoulos.
However, the outgoing code avoids the late time caustic formation noted in
Papadopoulos' work, as well as the complications of gauge ambiguity and
backscattering introduced by a finite outer boundary. One attractive option
would be to combine the best features of these approaches by matching an
interior evolution based upon ingoing null hypersurfaces to an exterior
evolution based upon outgoing null hypersurfaces, as implemented
in~\cite{luis2m} for spherically symmetric Einstein--Klein--Gordon waves.

The waveform of relevance to gravitational wave astronomy is the
superposition of modes with different frequency compositions and
angular dependence. Although this waveform results from a complicated
nonlinear processing of the input signal, which varies with choice of
observation angle, the response of the individual
modes to an input signal of arbitrary amplitude can be obtained by
scaling the response to an input of standard reference amplitude. This
offers an economical approach to preparing a waveform catalog.


\subsection{3D Einstein--Klein--Gordon system}
\label{sec:3dekg}

The Einstein--Klein--Gordon (EKG) system can be used to simulate many interesting
physical phenomena. In 1D, characteristic EKG codes have been used to simulate
critical phenomena and the perturbation of black holes (see Section~\ref{sec:1d}),
and a Cauchy EKG code has been used to study boson star
dynamics~\cite{SeidelSuen}. (The characteristic approach has not yet been
applied to the problem of stable 1D boson stars.) Extending these codes to 3D
would open up a new range of possibilities, e.g., the possibility to study
radiation from a boson star orbiting a black hole. A first step in that
direction has been achieved with the construction of a 3D characteristic code
by incorporating a massless scalar field into the PITT code~\cite{barretoekg}.
Since the scalar and gravitational evolution equations have the same basic
form, the same evolution algorithm could be utilized. The code was tested to be
second order convergent and stable. It was applied to the fully nonlinear
simulation of an asymmetric pulse of ingoing scalar radiation propagating
toward a Schwarzschild black hole. The resulting scalar radiation and
gravitational news backscattered to $\mathcal{I}^+$ was computed. The amplitudes
of the scalar and gravitational radiation modes exhibited the expected power
law scaling with respect to the initial pulse amplitude. In addition, the
computed ringdown frequencies agreed with the results from perturbative
quasinormal mode calculations.

\newpage


\section{Cauchy-Characteristic Matching}
\label{sec:ccm}

Characteristic evolution has many advantages over Cauchy evolution. Its one
disadvantage is the existence of either a caustic, where neighboring
characteristics focus, or a milder version consisting of a crossover between
two distinct characteristics. The vertex of a light cone is a highly symmetric
caustic which already strongly limits the time step for characteristic
evolution because of the CFL condition~(\ref{eq:cfl0}). It does not appear
possible for a single characteristic coordinate system to cover the entire
exterior region of a binary black hole spacetime without developing very
complicated caustics and crossovers. This limits the waveform determined by a
purely characteristic evolution to the post merger period.

CCM is a way to avoid such
limitations by combining the strong points of characteristic and
Cauchy evolution into a global evolution~\cite{Bis2}. One of the prime
goals of computational relativity is the simulation of the inspiral
and merger of binary black holes. Given the appropriate worldtube data
for a binary system in its interior, characteristic evolution can
supply the exterior spacetime and the radiated waveform. But
determination of the worldtube data for a binary requires an interior
Cauchy evolution. CCM is designed to solve such global problems. The
potential advantages of CCM over traditional boundary conditions are
\begin{itemize}
\item accurate waveform and polarization state at infinity,
\item computational efficiency for radiation problems in terms of both
  the grid domain and the computational algorithm,
\item elimination of an artificial outer boundary condition on the
  Cauchy problem, which eliminates contamination from back-reflection
  and clarifies the global initial value problem, and
\item a global picture of the spacetime exterior to the horizon.
\end{itemize}

These advantages have been realized in model tests, but CCM has not yet been
achieved in fully nonlinear three-dimensional general relativity. The early
attempts to implement CCM in general relativity involved the
Arnowitt--Deser--Misner (ADM)~\cite{adm} formulation for the Cauchy evolution.
The difficulties were later traced to a pathology in the way boundary
conditions have traditionally been applied in ADM codes. Instabilities
introduced at boundaries have emerged as a major problem common to all ADM code
development. A linearized study~\cite{belath,cauchboun} of ADM
evolution-boundary algorithms with prescribed values of lapse and
shift shows the following:
\begin{itemize}
\item On analytic grounds, ADM boundary algorithms which supply values
  for all components of the metric (or extrinsic curvature) are
  inconsistent.
\item A consistent boundary algorithm only allows free specification
  of the transverse-traceless components of the metric with respect to
  the boundary.
\item Using such a boundary algorithm, linearized ADM evolution can be
  carried out in a bounded domain for thousands of crossing times with
  no sign of exponential growing instability.
\end{itemize}

The evolution satisfied the original criterion for robust
stability~\cite{cauchboun}: that there be no exponential growth when the
initial Cauchy data and free boundary data are prescribed as random numbers (in
the linearized regime). These results gave some initial optimism that CCM might
be possible with an ADM code if the boundary condition was properly treated.
However, it was subsequently shown that ADM is only weakly hyperbolic so that
in the linear regime there are instabilities which grow as a power law in time.
In the nonlinear regime, it is symptomatic of weakly hyperbolic systems that
such secular instabilities become exponential. This has led to a refined
criterion for robust stability as a standardized test~\cite{apples}.

CCM cannot work unless the Cauchy and characteristic codes have robustly stable
boundaries. This is necessarily so because interpolations continually introduce
short wavelength noise into the neighborhood of the boundary. It was
demonstrated some time ago that the PITT characteristic code has a robustly
stable boundary (see Section~\ref{sec:stability}), but robustness of the Cauchy
boundary has only recently been studied.


\subsection{Computational boundaries}

Boundary conditions are both the most important and the most difficult
part of a theoretical treatment of most physical systems. Usually,
that's where all the physics is. And, in computational approaches,
that's usually where all the agony is. Computational boundaries for
hyperbolic systems pose special difficulties. Even with an analytic
form of the correct physical boundary condition in hand, there are
seemingly infinitely more unstable numerical implementations than
stable ones. In general, a stable problem places more boundary
requirements on the finite difference equations than on the
corresponding partial differential equations. Furthermore, the methods
of linear stability analysis are often more unwieldy to apply to the
boundary than to the interior evolution algorithm.

The von Neumann stability analysis of the interior algorithm linearizes the
equations, while assuming a uniform infinite grid, and checks that the discrete
Fourier modes do not grow exponentially. There is an additional stability
condition that a boundary introduces into this analysis. Consider the
one-dimensional case. The mode $e^{kx}$, with $k$ real, is not included in the
von Neumann analysis for periodic boundary conditions. However, for the half
plane problem with a boundary to the right on the $x$-axis, one can
legitimately prescribe such a mode  as initial data as long as $k>0$ so that it
has finite energy. Thus the stability of such boundary modes must  be checked.
In the case of an additional boundary to the left, the Godunov--Ryaben'kii
theory gives as necessary conditions for stability the separate von Neumann
stability of the interior and the stability of the allowed boundary
modes~\cite{sod}. The Kreiss condition~\cite{kreiss} strengthens this result
by providing a sufficient condition for stability.

The correct physical formulation of any asymptotically flat Cauchy problem also
involves asymptotic conditions at infinity. These conditions must ensure not
only that the total energy and the energy loss by radiation are both finite,
but they must also ensure the proper $1/r$ asymptotic falloff of the radiation
fields. However, when treating radiative systems computationally, an outer
boundary is often established artificially at some large but finite distance in
the wave zone, i.e.\ many wavelengths from the source. Imposing an appropriate
radiation boundary condition at a finite distance is a difficult task even in
the case of a simple radiative system evolving on a fixed geometric background.
The problem is exacerbated when dealing with Einstein's equation.

Nowhere is the boundary problem more acute than in the computation of
gravitational radiation produced by black holes. The numerical study of a black
hole spacetime by means of a pure Cauchy evolution involves inner as well as
outer grid boundaries. The inner boundary is necessary to avoid the
topological complications and singularities introduced by the black holes. For
multiple black holes, the inner boundary consists of disjoint pieces. Unruh
suggested the commonly accepted strategy for Cauchy evolution of black
holes (see~\cite{thornburg1987}). An inner boundary located at (or
near) an apparent horizon is used to excise the singular interior region.

CCM has a natural application to this problem. In
the Cauchy treatment of such a system, the outer grid boundary is located at
some finite distance, normally many wavelengths from the source. Attempts to
use compactified Cauchy hypersurfaces which extend to spatial infinity have
failed because the phase of short wavelength radiation varies rapidly in
spatial directions~\cite{Orsz}. Characteristic evolution avoids this problem
by approaching infinity along the phase fronts.

When the system is nonlinear and not amenable to an exact solution, a finite
outer boundary condition must necessarily introduce spurious physical effects
into a Cauchy evolution. The domain of dependence of the initial Cauchy data in
the region spanned by the computational grid would shrink in time along ingoing
characteristics unless data on a worldtube traced out by the outer grid
boundary is included as part of the problem. In order to maintain a causally
sensible evolution, this worldtube data must correctly substitute for the
missing Cauchy data which would have been supplied if the Cauchy hypersurface
had extended to infinity. In a scattering problem, this missing exterior Cauchy
data might, for instance, correspond to an incoming pulse initially outside the
outer boundary. In a problem where the initial radiation fields are confined to
a compact region inside the boundary, these missing Cauchy data are easy to
characterize when dealing with a constraint free field, such as a scalar field
$\Phi$ where the appropriate Cauchy data outside the boundary would be
$\Phi_{,t}=0$. However, the determination of Cauchy data for general relativity
is a global elliptic constraint problem so that there is no well defined scheme
to confine it to a compact region. Furthermore, even if the data for a given
problem were known on a complete initial hypersurface extending to infinity, it
would be a formidable nonlinear evolution problem to correctly assign the
associated boundary data on a finite outer boundary.

Another important issue arising in general relativity is whether the boundary
condition preserves the constraints. It is typical of hyperbolic reductions of
the Einstein equations that the Hamiltonian and momentum constraints propagate
in a domain of dependence dictated by the  light rays. Unless the boundary
conditions on the outer world tube enforce these constraints, they will be
violated outside the domain of dependence of the initial Cauchy hypersurface.
This issue of a constraint-preserving initial boundary value problem has only
recently been addressed~\cite{stewartbc}. The first fully nonlinear treatment
of a well-posed constraint preserving formulation of the Einstein
initial-boundary value problem (IBVP) has only recently been given by Friedrich and
Nagy~\cite{friednag}. Their treatment is based upon a frame formulation in
which the evolution variables are the tetrad, connection coefficients, and Weyl
curvature. Although this system has not yet been implemented computationally,
it has spurred the investigation of simpler treatments of Einstein
equations which give rise to a constraint preserving IBVP under
various restrictions~\cite{lehnercp,harm,reulacp,frittellicp,gundlachcp}.

It is common practice in computational physics to impose some artificial
boundary condition (ABC), such as an outgoing radiation condition, in an
attempt to approximate the proper data for the exterior region. This ABC may
cause partial reflection of an outgoing wave back into the
system~\cite{Lind,Orsz,Hig86,Ren}, which contaminates the accuracy of the
interior evolution and the calculation of the radiated waveform. Furthermore,
nonlinear waves intrinsically backscatter, which makes it incorrect to try to
entirely eliminate incoming radiation from the outer region. The resulting
error is of an analytic origin, essentially independent of computational
discretization. In general, a systematic reduction of this error can only be
achieved by moving the computational boundary to larger and larger radii. This
is computationally very expensive, especially for three-dimensional
simulations.

A traditional ABC for the wave equation is the
Sommerfeld condition. For a 3D scalar field this takes the form
$g_{,t}+g_{,r}=0$, where $g=r\Phi$. This condition is \emph{exact} only for
a linear wave with spherically symmetric data and boundary. In that
case, the exact solution is $g=f_1(t-r)+f_2(t+r)$ and the Sommerfeld
condition eliminates the incoming wave $f_2$.

Much work has been done on formulating boundary conditions, both exact and
approximate, for linear problems in situations that are not spherically
symmetric and in which the Sommerfeld condition would be inaccurate. These
boundary conditions are given various names in the literature, e.g., absorbing
or non-reflecting. A variety of ABC's have been reported for linear problems.
See the articles~\cite{giv,Ren,tsy,ryab,jcp97} for general discussions.

Local ABC's have been extensively applied to linear problems with varying
success~\cite{Lind,Eng77,Bay80,Tre86,Hig86,Bla88,Jia90}. Some of these
conditions are local approximations to exact integral representations of the
solution in the exterior of the computational domain~\cite{Eng77}, while others
are based on approximating the dispersion relation of the so-called one-way
wave equations~\cite{Lind,Tre86}. Higdon~\cite{Hig86} showed that this last
approach is essentially equivalent to specifying a finite number of angles of
incidence for which the ABC's yield perfect transmission. Local ABC's have
also been derived for the linear wave equation by considering the asymptotic
behavior of outgoing solutions~\cite{Bay80}, which generalizes the Sommerfeld
outgoing radiation condition. Although this type of ABC is relatively simple to
implement and has a low computational cost, the final accuracy is often limited
because the assumptions made about the behavior of the waves are rarely met in
practice~\cite{giv,tsy}.

The disadvantages of local ABC's have led some workers to consider
exact nonlocal boundary conditions based on integral representations of
the infinite domain problem~\cite{Tin86,giv,tsy}. Even for problems
where the Green's function is known and easily computed, such
approaches were initially dismissed as impractical~\cite{Eng77};
however, the rapid increase in computer power has made it possible to
implement exact nonlocal ABC's for the linear wave equation and
Maxwell's equations in 3D~\cite{deM,kell}. If properly implemented,
this kind of method can yield numerical solutions which converge to the
exact infinite domain problem in the continuum limit, while keeping the
artificial boundary at a fixed distance. However, due to nonlocality,
the computational cost per time step usually grows at a higher power with
grid size ($\mathcal{O} (N^4)$ per time step in three dimensions) than in a local
approach~\cite{giv,deM,tsy}.

The extension of ABC's to \emph{nonlinear} problems is much more difficult.
The problem is normally treated by linearizing the region between the
outer boundary and infinity, using either local or nonlocal linear
ABC's~\cite{tsy,ryab}. The neglect of the nonlinear terms in this
region introduces an unavoidable error at the analytic level. But even
larger errors are typically introduced in prescribing the outer
boundary data. This is a subtle global problem because the correct
boundary data must correspond to the continuity of fields and their
normal derivatives when extended across the boundary into the
linearized exterior. This is a clear requirement for any consistent
boundary algorithm, since discontinuities in the field or its
derivatives would otherwise act as spurious sheet source on the
boundary, which contaminates both the interior and the exterior
evolutions. But the fields and their normal derivatives constitute an
overdetermined set of data for the boundary problem. So it
is necessary to solve a global linearized problem, not just an exterior
one, in order to find the proper data. The designation ``exact ABC'' is
given to an ABC for a nonlinear system whose only error is due to
linearization of the exterior. An exact ABC requires the use of global
techniques, such as the difference potential method, to eliminate back
reflection at the boundary~\cite{tsy}.

To date there have been only a few applications of ABC's to strongly
nonlinear problems~\cite{giv}. Thompson~\cite{Tho87} generalized a
previous nonlinear ABC of Hedstrom~\cite{Hed79} to treat 1D and 2D
problems in gas dynamics. These boundary conditions performed poorly in
some situations because of their difficulty in adequately modeling the
field outside the computational domain~\cite{Tho87,giv}. Hagstrom and
Hariharan~\cite{Hag88} have overcome these difficulties in 1D gas
dynamics by a clever use of Riemann invariants. They proposed a
heuristic generalization of their local ABC to 3D, but this approach
has not yet been validated.

In order to reduce the level of approximation at the analytic level,
an artificial boundary for a nonlinear problem must be placed
sufficiently far from the strong-field region. This sharply increases
the computational cost in multi-dimensional simulations~\cite{Eng77}.
There seems to be no numerical method which converges (as the
discretization is refined) to the infinite domain exact solution of a
strongly nonlinear wave problem in multi-dimensions, while keeping the
artificial boundary fixed.

CCM is a strategy that eliminates this nonlinear
source of error. In the simplest version of CCM, Cauchy and characteristic
evolution algorithms are pasted together in the neighborhood of a worldtube to
form a global evolution algorithm. The characteristic algorithm provides an
\emph{outer} boundary condition for the interior Cauchy evolution, while the
Cauchy algorithm supplies an \emph{inner} boundary condition for the
characteristic evolution. The matching worldtube provides the geometric
framework necessary to relate the two evolutions. The Cauchy foliation slices
the worldtube into spherical cross-sections. The characteristic evolution is
based upon the outgoing null hypersurfaces emanating from these slices, with
the evolution proceeding from one hypersurface to the next by the outward
radial march described earlier. There is no need to truncate spacetime at a
finite distance from the sources, since compactification of the radial null
coordinate used in the characteristic evolution makes it possible to cover the
infinite space with a finite computational grid. In this way, the true
waveform may be directly computed by a finite difference algorithm. Although
characteristic evolution has limitations in the interior region where caustics
develop, it proves to be both accurate and computationally efficient in the
treatment of exterior regions.


\subsection{The computational matching strategy}
\label{sec:3dccm}

In its simplest form, CCM evolves a mixed spacelike-null initial value problem
in which Cauchy data is given in a spacelike region bounded by a spherical
boundary $\mathcal{S}$ and characteristic data is given on a null hypersurface
emanating from $\mathcal{S}$. The general idea is not entirely new. An early
mathematical investigation combining space-like and characteristic hypersurfaces
appears in the work of Duff~\cite{Duff}. The three chief ingredients for
computational implementation are: (i) a Cauchy evolution module, (ii) a
characteristic evolution module and, (iii) a module for matching the Cauchy and
characteristic regions across their interface. The interface is the timelike
worldtube which is traced out by the flow of $\mathcal{S}$ along the worldlines of
the Cauchy evolution, as determined by the choice of lapse and shift. Matching
provides the exchange of data across the worldtube to allow evolution without
any further boundary conditions, as would be necessary in either a purely Cauchy
or purely characteristic evolution. Other versions of CCM involve a finite
overlap between the characteristic and Cauchy regions.

The most important application of CCM is anticipated to be the binary black
hole problem. The 3D Cauchy codes being developed to solve this problem employ
a single Cartesian coordinate patch, a stategy adopted in~\cite{Alliance97b} to
avoid coordinate singularites. A thoroughly tested and robust 3D characteristic
code is now in place~\cite{high}, ready to match to the boundary of this Cauchy
patch. Development of a stable implementation of CCM represents the major step
necessary to provide a global evolution code for the binary black hole problem.

From a cursory view, the application of CCM to this problem might seem
routine, tantamount to translating into finite difference form the
textbook construction of an atlas consisting of overlapping coordinate
patches. In practice, it is an enormous project. The computational
strategy has
been outlined in~\cite{vishu}. The underlying geometrical algorithm
consists of the following main submodules:
\begin{itemize}
\item The \emph{boundary module} which sets the grid structures. This
  defines masks identifying which points in the Cauchy grid are to be
  evolved by the Cauchy module and which points are to be interpolated
  from the characteristic grid, and vice versa. The reference
  structures for constructing the mask is the inner characteristic
  boundary, which in Cartesian coordinates is the ``Euclidean''
  spherical worldtube $x^2+y^2+z^2=R^2$, and the outer Cauchy
  boundary. The choice of lapse and shift for the Cauchy evolution
  governs the dynamical and geometrical properties of the matching
  worldtube.
\item The \emph{extraction module} whose input is Cauchy grid data in
  the neighborhood of the worldtube and whose output is the inner
  boundary data for the exterior characteristic evolution. This module
  numerically implements the transformation from Cartesian \{3\,+\,1\}
  coordinates to spherical null coordinates. The algorithm makes no
  perturbative assumptions and is based upon interpolations of the
  Cauchy data to a set of prescribed grid points on the worldtube. The
  metric information is then used to solve for the null geodesics
  normal to the slices of the worldtube. This provides the Jacobian
  for the transformation to null coordinates in the neighborhood of
  the worldtube. The characteristic evolution module is then used to
  propagate the data from the worldtube to null infinity, where the
  waveform is calculated.
\item The \emph{injection module} which completes the interface by
  using the exterior characteristic evolution to supply the outer
  boundary data for the Cauchy evolution. This is the inverse of the
  extraction procedure but must be implemented outside the worldtube
  to allow for overlap between Cauchy and characteristic domains. The
  overlap region can be constructed either to have a fixed physical
  size or to shrink to zero in the continuum limit. In the latter
  case, the inverse Jacobian describing the transformation from null
  to Cauchy coordinates can be obtained to prescribed accuracy in
  terms of an affine parameter expansion along the null geodesics
  emanating from the worldtube. But the numerical stability of the
  scheme is not guaranteed.
\end{itemize}

The above strategy provides a model of how Cauchy and characteristic codes
can be pieced together as modules to form a global evolution code.

The full advantage of CCM lies in the numerical treatment of
nonlinear systems where its error converges to zero in the continuum
limit of infinite grid resolution~\cite{Bis,Bis2,Clarke}. For high
accuracy, CCM is also by far the most efficient method. For small
target error $\varepsilon$, it has been shown that the relative
amount of computation required for CCM ($A_\mathrm{CCM}$) compared to that
required for a pure Cauchy calculation ($A_\mathrm{C}$) goes to zero,
$A_\mathrm{CCM}/A_\mathrm{C} \rightarrow O$ as
$\varepsilon \rightarrow O$~\cite{cce,vishu}. An important factor here
is the use of a compactified characteristic evolution, so that the
whole spacetime is represented on a finite grid. From a numerical
point of view this means that the only error made in a calculation of
the radiation waveform at infinity is the controlled error due to the
finite discretization. Accuracy of a Cauchy algorithm which uses an
ABC requires a large grid domain in order to avoid error from
nonlinear effects in its exterior. The computational demands of
CCM are small because the interface problem involves one less
dimension than the evolution problem. Because characteristic evolution
algorithms are more efficient than Cauchy algorithms, the efficiency
can be further enhanced by making the matching radius as small as possible
consistent with the avoidance of caustics.

At present, the computational strategy of CCM is exclusively the tool of
general relativists who are used to dealing with novel coordinate systems. A
discussion of its potential is given in~\cite{Bis}. Only
recently~\cite{Clarke,cylinder1,cylinder2,Ccprl,harm} has its practicability been
carefully explored. Research on this topic has been stimulated by the
requirements of the Binary Black Hole Grand Challenge Alliance, where CCM was
one of the strategies pursued to provide boundary conditions and
determine the radiation waveform. But I anticipate that its use will eventually
spread throughout computational physics because of its inherent advantages in
dealing with hyperbolic systems, particularly in three-dimensional problems
where efficiency is desired. A detailed study of the stability and accuracy of
CCM for linear and nonlinear wave equations has been presented
in~\cite{jcp97}, illustrating its potential for a wide range of
problems.


\subsection{Perturbative matching schemes}

In numerous analytic applications outside of general relativity,
matching techniques have successfully cured pathologies in perturbative
expansions~\cite{aliney}. Matching is a strategy for obtaining a global
solution by patching together solutions obtained using different
coordinate systems for different regions. By adopting each coordinate
system to a length scale appropriate to its domain, a globally
convergent perturbation expansion is sometimes possible in cases where
a single coordinate system would fail.

In general relativity, Burke showed that matching could be used to eliminate
some of the divergences arising in perturbative calculations of gravitational
radiation~\cite{burke}. Kates and Kegles further showed that use of an exterior
null coordinate system in the matching scheme could eliminate problems in the
perturbative treatment of a scalar radiation field on a Schwarzschild
background~\cite{kk}. The Schwarzschild light cones have drastically different
asymptotic behavior from the artificial Minkowski light cones used in
perturbative expansions based upon a flat space Green function. Use of the
Minkowski light cones leads to \emph{nonuniformities} in the expansion of the
radiation fields which are eliminated by the use of true null coordinates in
the exterior. Kates, Anderson, Kegles, and Madonna extended this work to the
fully general relativistic case and reached the same conclusion~\cite{kakm}.
Anderson later applied this approach to the slow motion approximation of a
binary system and obtained a derivation of the radiation reaction effect on the
orbital period which avoided some objections to earlier
approaches~\cite{and87}. The use of the true light cones was also essential in
formulating as a mathematical theorem that the Bondi news function satisfies
the Einstein quadrupole formula to leading order in a Newtonian
limit~\cite{quad}. Although questions of mathematical consistency still remain
in the perturbative treatment of gravitational radiation, it is clear that the
use of characteristic methods pushes these problems to a higher perturbative
order.

One of the first applications of characteristic matching was a hybrid
numerical-analytical treatment by Anderson and Hobill of the test problem of
nonlinear 1D scalar waves~\cite{Hobill,Hobill2,Hobill3}. They matched an inner
numerical solution to a  far field solution which was obtained by a
perturbation expansion. A key ingredient is that the far field is solved in
retarded null coordinates $(u,r)$. Because the transformation from null
coordinates $(u,r)$ to Cauchy coordinates $(t,r)$ is known analytically for
this problem, the matching between the null and Cauchy solutions is quite
simple. Causality was enforced by requiring that the system be stationary prior
to some fixed time. This eliminates extraneous incoming radiation in a
physically correct way in a system which is stationary prior to a fixed time
but it is nontrivial to generalize, say, to the problem of radiation from an
orbiting binary.

Later, a global, characteristic, numerical study of the self-gravitating
version of this problem, by G\'{o}mez and Winicour, confirmed that the use of
the true null cones is essential in getting the correct radiated
waveform~\cite{finw}. For quasi-periodic radiation, the phase of the
waveform is
particular sensitive to the truncation of the outer region at a finite
boundary. Although a perturbative estimate would indicate an $\mathcal{O} (M/R)$ error,
this error accumulates over many cycles to produce an error of order $\pi$ in
the phase.

Anderson and Hobill proposed that their method be extended to general
relativity by matching a numerical solution to an analytic $1/r$ expansion in
null coordinates. Most perturbative-numerical matching schemes that have been
implemented in general relativity have been based upon perturbations of a
Schwarzschild background using the standard Schwarzschild time
slicing~\cite{ab1,ab2,ab3,all1,rupright,rezzmatz,nagar}. It would be
interesting to compare results with an analytic-numeric matching scheme based
upon the true null cones. Although the full proposal by Anderson and Hobill
has not been carried out, characteristic techniques have been
used~\cite{lous,zlochadv,zlochret} to study the radiation content of numerical
solutions by treating the far field as a perturbation of a Schwarzschild
spacetime.

Most metric based treatments of gravitational radiation based upon
perturbations of Schwarzschild solve the underlying Regge--Wheeler~\cite{regge}
and Zerilli~\cite{zerilli} equations using traditional spacelike Cauchy
hypersurfaces. At one level, these approaches \emph{extract} the radiation from
a numerical solution in a region with outer boundary $\mathcal{B}$ by using data
on an inner worldtube $\mathcal{W}$ to construct the perturbative solution.
Ambiguities are avoided by use of gauge invariant perturbation
quantities~\cite{moncrief}. For this to work, $\mathcal{W}$  must not only be
located in the far field but, because of the lack of proper outer boundary
data, it is necessary that the boundary $\mathcal{B}$  be sufficiently far outside
$\mathcal{W}$ so that the extracted radiation is uncontaminated by back-reflection
for some significant window of time. This poses extreme computational
requirements in a 3D problem. This extraction strategy has also been carried
out using characteristic evolution in the exterior of $\mathcal{W}$ instead of a
perturbative solution, i.e.\ Cauchy-characteristic extraction~\cite{cce}.
Babiuc, Szil{\'{a}}gyi, Hawke, and Zlochower have recently carried out a test
comparison of the two methods~\cite{babiuc}.

The contamination of the extracted radiation by back-reflection can only be
eliminated by matching to an exterior solution which \emph{injects} the
physically appropriate boundary data on $\mathcal{W}$. Cauchy-perturbative
matching~\cite{rupright,rezzmatz} has been implemented using the same modular
structure described for CCM in Section~\ref{sec:3dccm}. Nagar and
Rezzolla~\cite{nagar} have given a  review of this approach. At present,
perturbative matching and CCM share the common problem of long term stability
of the outer Cauchy boundary in 3D applications.


\subsection{Cauchy-characteristic matching for 1D gravitational systems}

The first numerical implementations of CCM were 1D feasibility studies.
These model problems provided a controlled environment for the
development of CCM, in which either exact solutions or independent numerical
solutions were known. The following studies showed that CCM worked like
a charm in a variety of 1D applications, i.e.\ the matched evolutions were
essentially transparent to the presence of the interface.


\subsubsection{Cylindrical matching}
\label{sec:cylmatch}

The Southampton group chose cylindrically symmetric systems as their model
problem for developing matching techniques. In preliminary work, they showed
how CCM could be consistently carried out for a scalar wave evolving in
Minkowski spacetime but expressed in a nontrivial cylindrical coordinate
system~\cite{Clarke}.

They then tackled the gravitational problem. First they set up the analytic
machinery necessary for investigating cylindrically symmetric vacuum
spacetimes~\cite{cylinder1}. Although the problem involves only one spatial
dimension, there are two independent modes of polarization. The Cauchy metric
was treated in the Jordan--Ehlers--Kompaneets canonical form, using coordinates
$(t,r,\phi,z)$ adapted to the $(\phi,z)$ cylindrical symmetry. The advantage
here is that $u=t-r$ is then a null coordinate which can be used for the
characteristic evolution. They successfully recast the equations in a suitably
regularized form for the compactification of $\mathcal{I}^+$ in terms of the
coordinate $y=\sqrt{1/r}$. The simple analytic relationship between Cauchy
coordinates $(t,r)$ and characteristic coordinates $(u,y)$ facilitated the
translation between Cauchy and characteristic variables on the matching
worldtube, given by $r = \mathrm{const}$.

Next they implemented the scheme as a numerical code. The interior Cauchy
evolution was carried out using an unconstrained leapfrog scheme. It is notable
that they report no problems with instability, which have arisen in other
attempts at unconstrained leapfrog evolution in general relativity. The
characteristic evolution also used a leapfrog scheme for the evolution between
retarded time levels $u$, while numerically integrating the hypersurface
equations outward along the characteristics.

The matching interface was located at points common to both the Cauchy and
characteristic grids. In order to update these points by Cauchy evolution, it
was necessary to obtain field values at the Cauchy ``guard'' points which lie
outside the worldtube in the characteristic region. These values were obtained
by interpolation from characteristic grid points (lying on three levels of null
hypersurfaces in order to ensure second order accuracy). Similarly, the
boundary data for starting up the characteristic integration was obtained by
interpolation from Cauchy grid values inside the worldtube.

The matching code was first tested~\cite{cylinder2} using exact Weber--Wheeler
cylindrical waves~\cite{wweb}, which come in from $\mathcal{I}^-$, pass through
the symmetry axis and expand out to $\mathcal{I}^+$. The numerical errors were
oscillatory with low growth rate, and second order convergence was confirmed.
Of special importance, little numerical noise was introduced by the interface.
Comparisons of CCM  were made with Cauchy evolutions using a standard outgoing
radiation boundary condition~\cite{piran}. At high amplitudes the standard
condition developed a large error very quickly and was competitive only for
weak waves with a large outer boundary. In contrast, the matching code
performed well even with a small matching radius. Some interesting simulations
were presented in which an outgoing wave in one polarization mode collided with
an incoming wave in the other mode, a problem studied earlier by pure Cauchy
evolution~\cite{stark}. The simulations of the collision were qualitatively
similar in these two studies.

The Weber--Wheeler waves contain only one gravitational degree of freedom. The
code was next tested~\cite{south3} using exact cylindrically symmetric
solutions, due to Piran, Safier, and Katz~\cite{katz}, which contain both
degrees of freedom. These solutions are singular at $\mathcal{I}^+$ so that the
code had to be suitably modified. Relative errors of the various metric
quantities were in the range $10^{-4}$ to $10^{-2}$. The convergence rate of
the numerical solution starts off as second order but diminishes to first order
after long time evolution. This performance could perhaps be improved by
incorporating subsequent improvements in the characteristic code made by
Sperhake, Sj\" odin, and Vickers (see Section~\ref{sec:1d}).


\subsubsection{Spherical matching}

A joint collaboration between groups at Pennsylvania State University and the
University of Pittsburgh applied CCM to the EKG
system with spherical symmetry~\cite{ekgmat}. This model problem
allowed simulation of black hole formation as well as wave
propagation.

The geometrical setup is analogous to the cylindrically symmetric problem.
Initial data were specified on the union of a spacelike hypersurface and a null
hypersurface. The evolution used a 3-level Cauchy scheme in the interior and a
2-level characteristic evolution in the compactified exterior. A constrained
Cauchy evolution was adopted because of its earlier success in accurately
simulating scalar wave collapse~\cite{choptprl}. Characteristic evolution was
based upon the null parallelogram algorithm~(\ref{eq:integral}). The
matching between the Cauchy and characteristic foliations was achieved by
imposing continuity conditions on the metric, extrinsic curvature and scalar
field variables, ensuring smoothness of fields and their derivatives across the
matching interface. The extensive analytical and numerical studies of this
system in recent years aided the development of CCM in this non-trivial
geometrical setting by providing basic knowledge of the
expected physical and geometrical behavior, in the absence of exact solutions.

The CCM code accurately handled wave propagation and
black hole formation for all values of $M/R$ at the matching radius,
with no symptoms of instability or back-reflection. Second order
accuracy was established by checking energy conservation.


\subsubsection{Excising 1D black holes}

In further developmental work on the EKG model, the Pittsburgh group used CCM
to formulate a new treatment of the inner Cauchy boundary for a black hole
spacetime~\cite{excise}. In the conventional approach, the inner boundary of
the Cauchy evolution is located at an apparent horizon, which must lie inside
(or on) the event horizon~\cite{wald1984}, so that truncation of the interior
spacetime at the apparent horizon cannot causally affect the gravitational
waves radiated to infinity. This is the physical rationale behind the apparent
horizon boundary condition. However, instabilities reported in some early
attempts at the conventional approach motivated an alternative treatment.

\begin{figure}[hptb]
  \def\epsfsize#1#2{0.6#1}
  \centerline{\epsfbox{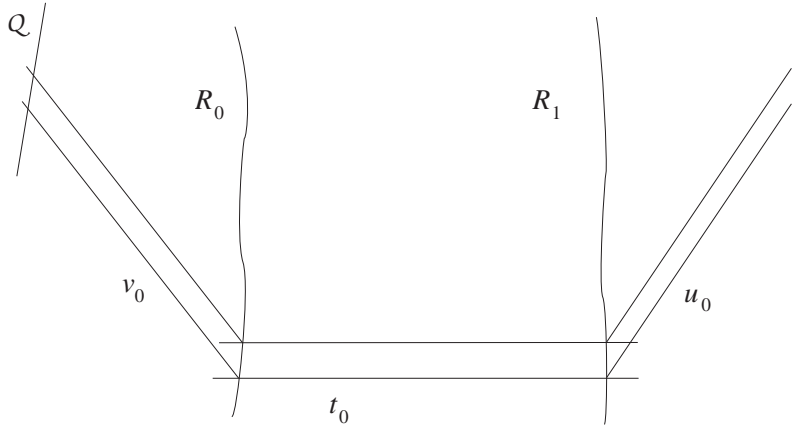}}
  \caption{\it Black hole excision by matching. A Cauchy evolution,
    with data at $t_0$ is matched across worldtubes $R_0$ and $R_1$
    to an ingoing null evolution, with data at $v_0$, and an
    outgoing null evolution, with data at $u_0$. The ingoing null
    evolution extends to an inner trapped boundary $Q$, and the
    outgoing null evolution extends to $\mathcal{I}^+$.}
  \label{fig:1dexci}
\end{figure}

In the CCM strategy, illustrated in Figure~\ref{fig:1dexci}, the interior black
hole region is evolved using an \emph{ingoing} null algorithm whose inner
boundary is a marginally trapped surface, and whose outer boundary lies outside
the black hole and forms the inner boundary of a region evolved by the Cauchy
algorithm. In turn, the outer boundary of the Cauchy region is handled by
matching to an outgoing null evolution extending to $\mathcal{I}^+$. Data are
passed between the inner characteristic and central Cauchy regions using a CCM
procedure similar to that already described for an outer Cauchy boundary. The
main difference is that, whereas the outer Cauchy boundary data is induced
from the Bondi metric on an outgoing null hypersurface, the inner Cauchy
boundary is now obtained from an ingoing null hypersurface which enters the
event horizon and terminates at a marginally trapped surface.

The translation from an outgoing to an incoming null evolution algorithm can be
easily carried out. The substitution $\beta\rightarrow \beta+i\pi/2$ in the 3D
version of the Bondi metric~(\ref{eq:metric}) provides a simple formal recipe
for switching from an outgoing to an ingoing null formalism~\cite{excise}.

In order to ensure that trapped surfaces exist on the ingoing
null hypersurfaces, initial data were chosen which guarantee black hole
formation. Such data can be obtained from initial Cauchy data for a
black hole. However, rather than extending the Cauchy hypersurface
inward to an apparent horizon, it was truncated sufficiently far outside
the apparent horizon to avoid computational problems with the Cauchy
evolution. The initial Cauchy data were then extended into the black hole
interior as initial null data until a marginally trapped surface was
reached. Two ingredients were essential in order to arrange this.
First, the inner matching surface must be chosen to be convex, in the
sense that its outward null normals uniformly diverge and its inner
null normals uniformly converge. (This is trivial to satisfy in the
spherically symmetric case.) Given any physically reasonable matter
source, the focusing theorem guarantees that the null rays
emanating inward from the matching sphere continue to converge until
reaching a caustic. Second, the initial null data must lead to a
trapped surface before such a caustic is encountered. This is a
relatively easy requirement to satisfy because the initial null data
can be posed freely, without any elliptic or algebraic constraints
other than continuity with the Cauchy data.

A code was developed which implemented CCM at both the inner and outer
boundaries~\cite{excise}. Its performance showed that CCM provides as good a
solution to the black hole excision problem in spherical symmetry as any
previous treatment~\cite{scheel1995a,scheel1995b,mc,anninos1995}. CCM is
computationally more efficient than these pure Cauchy approaches (fewer
variables) and much easier to implement. Depending upon the Cauchy formalism
adopted, achieving stability with a pure Cauchy scheme in the region of an
apparent horizon can be quite tricky, involving much trial and error in
choosing finite difference schemes. There were no complications with stability
of the null evolution at the marginally trapped surface.

The Cauchy evolution was carried out in ingoing Eddington--Finklestein
(IEF) coordinates. The initial Cauchy data consisted of a
Schwarzschild black hole with an ingoing Gaussian pulse of scalar
radiation. Since IEF coordinates are based on ingoing null cones, it
is possible to construct a simple transformation between the IEF Cauchy
metric and the ingoing null metric. Initially there was no scalar field
present on either the ingoing or outgoing null patches. The initial
values for the Bondi variables $\beta$ and $V$ were determined by
matching to the Cauchy data at the matching surfaces and integrating
the hypersurface equations~(\ref{eq:sbeta}, \ref{eq:sv}).

As the evolution proceeds, the scalar field passes into the black hole, and the
marginally trapped surface (MTS) grows outward. The MTS is easily located
in the spherically symmetric case by an algebraic equation. In order to excise
the singular region, the grid points inside the marginally trapped surface were
identified and masked out of the evolution. The backscattered radiation
propagated cleanly across the outer matching surface to $\mathcal{I}^+$. The
strategy worked smoothly, and second order accuracy of the approach was
established by comparing it to an independent numerical solution obtained using
a second order accurate, purely Cauchy code~\cite{mc}. As discussed in
Section~\ref{sec:bbhib}, this inside-outside application of CCM has potential
application to the binary black hole problem.

In a variant of this double CCM matching scheme, Lehner~\cite{luis2m} has
eliminated the middle Cauchy region and constructed a 1D code matching the
ingoing and outgoing characteristic evolutions directly across a single
timelike worldtube. In this way, he is able to simulate the global problem of a
scalar wave falling into a black hole by purely characteristic methods.


\subsection{Axisymmetric Cauchy-characteristic matching}
\label{sec:aximatch}

The Southampton CCM project is being carried out for spacetimes with (twisting)
axial symmetry. The formal basis for the matching scheme was developed by
d'Inverno and Vickers~\cite{south1,south2}. Similar to the Pittsburgh 3D
strategy (see Section~\ref{sec:3dccm}), matching is based upon an extraction
module, which supplies boundary data for the exterior characteristic evolution,
and an injection module, which supplies boundary data for the interior Cauchy
evolution. However, their use of spherical coordinates for the Cauchy evolution
(as opposed to Cartesian coordinates in the 3D strategy) allows use of a
matching worldtube $r=R_\mathrm{m}$ which lies simultaneously on Cauchy and
characteristic gridpoints. This tremendously simplifies the necessary
interpolations between the Cauchy and characteristic evolutions, at the
expense of dealing with the $r=0$ coordinate singularity in the Cauchy
evolution. The characteristic code (see Section~\ref{sec:axiev}) is based upon a
compactified Bondi--Sachs formalism. The use of a ``radial'' Cauchy gauge, in
which the Cauchy coordinate $r$ measures the surface area of spheres,
simplifies the relation to the Bondi--Sachs coordinates. In the numerical
scheme, the metric and its derivatives are passed between the Cauchy and
characteristic evolutions exactly at $r=R_\mathrm{m}$, thus eliminating the need of a
matching interface encompassing a few grid zones, as in the 3D Pittsburgh
scheme. This avoids a great deal of interpolation error and computational
complexity.

Preliminary results in the development of the Southampton CCM code are
described by Pollney in his thesis~\cite{pollney}. The Cauchy code is based
upon the axisymmetric ADM code of Stark and Piran~\cite{starkpir} and
reproduces their vacuum results for a short time period, after which an
instability at the origin becomes manifest. The characteristic code has been
tested to reproduce accurately the Schwarzschild and boost-rotation symmetric
solutions~\cite{boostrot}, with more thorough tests of stability and accuracy
still to be carried out.


\subsection{Cauchy-characteristic matching for 3D scalar waves}

CCM has been successfully implemented in the fully 3D problem of
nonlinear scalar waves evolving in a flat spacetime~\cite{jcp97,Ccprl}.
This study demonstrated the feasibility of
matching between Cartesian Cauchy coordinates and spherical null
coordinates, the setup required to apply CCM to the binary
black hole problem. Unlike spherically or cylindrically symmetric
examples of matching, the
Cauchy and characteristic patches do not share a common coordinate
which can be used to define the matching interface. This introduces a
major complication into the matching procedure, resulting in extensive
use of inter-grid interpolation. The accompanying short wavelength
numerical noise presents a new challenge in obtaining a stable
algorithm.

The nonlinear waves were modeled by the equation
\begin{equation}
  c^{-2} \partial_{tt} \Phi = \nabla^{2} \Phi + F (\Phi) + S (t,x,y,z),
  \label{eq:swe}
\end{equation}
with self-coupling $F(\Phi)$ and external source $S$. The initial
Cauchy data $\Phi(t_0,x,y,z)$ and $\partial_t\Phi(t_0,x,y,z)$ are
assigned in a spatial region bounded by a spherical matching surface of
radius $R_\mathrm{m}$.

The characteristic initial value problem~(\ref{eq:swe}) is expressed in
standard spherical coordinates $(r,\theta,\varphi)$ and retarded time
$u=t-r+R_\mathrm{m}$:
\begin{equation}
  2 \partial_{ur} g =
  \partial_{rr} g - \frac{L^2 g}{r^2} + r (F + S),
  \label{eq:SWE}
\end{equation}
where $g = r\Phi$ and $L^2$ is the angular momentum operator
\begin{equation}
  L^2 g = - \frac{\partial_\theta (\sin \theta \, \partial_{\theta} g)}
  {\sin \theta} - \frac{\partial_\varphi^2 g}{\sin^2 \theta}.
\end{equation}
The initial null data consist of $g(r,\theta,\varphi,u_0)$ on the outgoing
characteristic cone $u_0 =t_0$ emanating at the initial Cauchy time from the
matching worldtube at $r= R_\mathrm{m}$.

CCM was implemented so that, in the continuum limit, $\Phi$ and its normal
derivatives would be continuous across the matching interface. The use of a
Cartesian discretization in the interior and a spherical discretization in the
exterior complicated the treatment of the interface. In particular, the
stability of the matching algorithm required careful attention to the details
of the inter-grid matching. Nevertheless, there was a reasonably broad range
of discretization parameters for which CCM was stable.

Two different ways of handling the spherical coordinates were used. One was
based upon two overlapping stereographic grid patches and the other upon a
multiquadric approximation using a quasi-regular triangulation of the sphere.
Both methods gave similar accuracy. The multiquadric method showed a slightly
larger range of stability. Also, two separate tactics were used to implement
matching, one based upon straightforward interpolations and the other upon
maintaining continuity of derivatives in the outward null direction (a
generalization of the Sommerfeld condition). Both methods were stable for a
reasonable range of grid parameters. The solutions were second order accurate
and the Richardson extrapolation technique could be used to accelerate
convergence.

The performance of CCM was compared to traditional ABC's. As expected, the
nonlocal ABC's yielded convergent results only in linear problems, and
convergence was not observed for local ABC's, whose restrictive assumptions
were violated in all of the numerical experiments. The computational cost of
CCM was much lower than that of current nonlocal boundary conditions. In
strongly nonlinear problems, CCM appears to be the only available method which
is able to produce numerical solutions which converge to the exact solution
with a fixed boundary.


\subsection{Stable implementation of 3D linearized Cauchy-characteristic matching}
\label{sec:linccm}

Although the individual pieces of the CCM module have been calibrated to give
a second order accurate interface between Cauchy and characteristic evolution
modules in 3D general relativity, its stability has not yet been
established~\cite{vishu}. However, a stable version of CCM for linearized
gravitational theory has recently been demonstrated~\cite{harm}. The Cauchy
evolution is carried out using a harmonic formulation for which the reduced
equations have a well-posed initial-boundary problem. Previous attempts at CCM
were plagued by boundary induced instabilities of the Cauchy code. Although
stable behavior of the Cauchy boundary is only a necessary and not a
sufficient condition for CCM, the tests with the linearized harmonic code
matched to a linearized characteristic code were successful.

\begin{figure}[hptb]
  \def\epsfsize#1#2{1.0#1}
  \centerline{\epsfbox{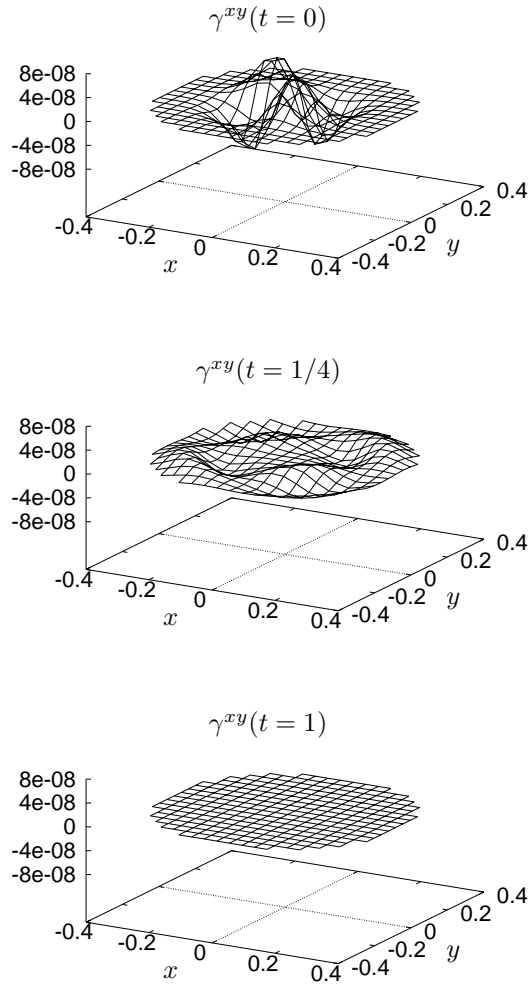}}
  \caption{\it Sequence of slices of the metric component
    $\gamma^{xy}$, evolved with the linear matched
    Cauchy-characteristic code. In the last snapshot, the wave has
    propagated cleanly onto the characteristic grid with negligible
    remnant noise.}
  \label{fig:2D.ccm}
\end{figure}

The harmonic conditions consist of wave equations which can be used to
propagate the gauge as four scalar waves using characteristic evolution. This
allows the extraction world tube to be placed at a finite distance from the
injection world tube without introducing a gauge ambiguity. Furthermore, the
harmonic gauge conditions are the only constraints on the Cauchy formalism so
that gauge propagation also insures constraint propagation. This allows the
Cauchy data to be supplied in numerically benign Sommerfeld form,  without
introducing constraint violation. Using random initial data, robust stability
of the CCM algorithm was confirmed for 2000 crossing times on a $45^3$ Cauchy
grid. Figure~\ref{fig:2D.ccm} shows a sequence of profiles  of the metric
component $\gamma^{xy}=\sqrt{-g}g^{xy}$ as a linearized wave propagates
cleanly through the spherical injection boundary and passes to the
characteristic grid, where it is propagated to $\mathcal{I}^+$.

\subsection{The binary black hole inner boundary}
\label{sec:bbhib}

It is clear that the three-dimensional inspiral and coalescence of black
holes challenges the limits of present computational know-how. CCM
offers a new approach for excising an interior trapped region which
might provide the enhanced flexibility required to solve this problem. In a
binary system, there are major computational advantages in posing the
Cauchy evolution in a frame which is co-rotating with the orbiting
black holes. Such a description seems necessary in order to keep the
numerical grid from being intrinsically twisted. In this
co-orbiting description, the Cauchy evolution requires an inner
boundary condition inside the black holes and also an outer boundary
condition on a worldtube outside of which the grid rotation is likely
to be superluminal. An outgoing characteristic code can routinely
handle such superluminal gauge flows in the exterior~\cite{high}. Thus,
successful implementation of CCM would solve the exterior boundary
problem for this co-orbiting description.

CCM also has the potential to handle the two black holes inside the Cauchy
region. As described earlier with respect to Figure~\ref{fig:1dexci}, an ingoing
characteristic code can evolve a moving black hole with long term
stability~\cite{excise,wobb}. This means that CCM might also be able
to provide the inner
boundary condition for Cauchy evolution once stable matching has been
accomplished. In this approach, the interior boundary of the Cauchy evolution
is located \emph{outside} the apparent horizon and matched to a characteristic
evolution based upon ingoing null cones. The inner boundary for the
characteristic evolution is a trapped or marginally trapped surface, whose
interior is excised from the evolution.

In addition to restricting the Cauchy evolution to the region outside the black
holes, this strategy offers several other advantages. Although finding a
marginally trapped surface on the ingoing null hypersurfaces remains an
elliptic problem, there is a natural radial coordinate system $(r,\theta,\phi)$
to facilitate its solution. Motion of the black hole through the grid reduces
to a one-dimensional radial problem, leaving the angular grid intact and thus
reducing the computational complexity of excising the inner singular region.
(The angular coordinates can even rotate relative to the Cauchy coordinates in
order to accommodate spinning black holes.) The chief danger in this approach
is that a caustic might be encountered on the ingoing null hypersurface before
entering the trapped region. This is a gauge problem whose solution lies in
choosing the right location and geometry of the surface across which the Cauchy
and characteristic evolutions are matched. There is a great deal of flexibility
here because the characteristic initial data can be posed without constraints.

\begin{figure}[hptb]
  \def\epsfsize#1#2{0.3#1}
  \centerline{\epsfbox{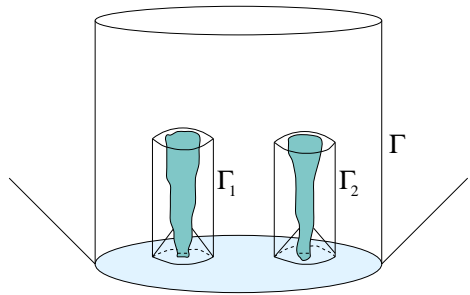}}
  \caption{\it CCM for binary black holes, portrayed in a
    co-rotating frame. The Cauchy evolution is matched across two
    inner worldtubes $\Gamma_1$ and $\Gamma_2$ to two ingoing null
    evolutions whose inner boundaries excise the individual black
    holes. The outer Cauchy boundary is matched across the worldtube
    $\Gamma$ to an outgoing null evolution extending to
    $\mathcal{I}^+$.}
  \label{fig:canbbh}
\end{figure}

This global strategy is tailor-made to treat two black holes in the co-orbiting
gauge, as illustrated in Figure~\ref{fig:canbbh}. Two disjoint characteristic
evolutions based upon ingoing null cones are matched across worldtubes to
a central Cauchy region. The interior boundaries of each of these interior
characteristic regions border a trapped surface. At the outer boundary of
the Cauchy region, a matched characteristic evolution based upon outgoing null
hypersurfaces propagates the radiation to infinity.

Present characteristic and Cauchy codes can handle the individual pieces of
this problem. Their unification appears to offer the best chance for
simulating the inspiral and merger of two black holes. The individual pieces
of the fully nonlinear CCM module, as outlined in Section~\ref{sec:3dccm}, have
been implemented and tested for accuracy. The one missing ingredient is long
term stability in the nonlinear gravitational case, which would make future
applications very exciting.

\newpage


\section{Numerical Hydrodynamics on Null Cones}
\label{sec:grace}

Numerical evolution of relativistic hydrodynamics has been traditionally
carried out on spacelike Cauchy hypersurfaces. Although the Bondi--Sachs
evolution algorithm can easily be extended to include matter~\cite{isaac}, the
advantage of a light cone approach for treating fluids is not as apparent as
for a massless field whose physical characteristics lie on the light cone.
However, results from recent studies of  relativistic stars and of fluid
sources moving in the vicinity of a black hole indicate that this approach can
provide accurate simulations of  astrophysical relevance such as supernova
collapse to a black hole, mass accretion, and the production of gravitational
waves.


\subsection{Spherically symmetric hydrodynamic codes}
\label{sec:shydro}

The earliest fully general relativistic simulations of fluids were carried out
in spherical symmetry. The first major work was a study of gravitational
collapse by May and White~\cite{may}. Most of the early work was carried out
using Cauchy evolution~\cite{tonirev}. Miller and Mota~\cite{miller} performed
the first simulations of spherically symmetric gravitational collapse using a
null foliation. Baumgarte, Shapiro and Teukolsky subsequently used a null
slicing to study supernovae~\cite{bst1} and the collapse of neutron stars to
form black holes~\cite{bst2}. The use of a null slicing allowed them to evolve
the exterior spacetime while avoiding the region of singularity formation.

The group at the Universidad de Oriente in Venezuela applied characteristic
methods to study the self-similar collapse of spherical matter and charge
distributions~\cite{barreto96,barreto98,barreto99}. The assumption of
self-similarity reduces the problem to a system of ODE's, subject to boundary
conditions determined by matching to an exterior Reissner--Nordstr\"om--Vaidya
solution. Heat flow in the internal fluid is balanced at the surface by the
Vaidya radiation. Their simulations illustrate how a nonzero total charge can
halt gravitational collapse and produce a final stable
equilibrium~\cite{barreto99}. It is interesting that the pressure vanishes in
the final equilibrium state so that hydrostatic support is completely supplied
by Coulomb repulsion.

Font and Papadopoulos~\cite{toniphi} have given a state-of-the-art treatment
of relativistic fluids which is applicable to either spacelike or null
foliations. Their approach is based upon a high-resolution shock-capturing
(HRSC) version of relativistic hydrodynamics in flux conservative form, which
was developed by the Valencia group (for a review see~\cite{tonirev}).
In the HRSC scheme, the hydrodynamic equations are written in flux
conservative, hyperbolic form. In each computational cell, the system of
equations is diagonalized to determine the characteristic fields and
velocities, and the local Riemann problem is solved to obtain a solution
consistent with physical discontinuities. This allows a finite differencing
scheme along the characteristics of the fluid that preserves the conserved
physical quantities and leads to a stable and accurate treatment of shocks.
Because the general relativistic system of hydrodynamical equations is
formulated in covariant form, it can equally well be applied to spacelike or
null foliations of the spacetime. The null formulation gave remarkable
performance in the standard Riemann shock tube test carried out in a Minkowski
background. The code was successfully implemented first in the case of
spherical symmetry, using a version of the Bondi--Sachs formalism adapted to
describe gravity coupled to matter with a worldtube boundary~\cite{tam}. They
verified second order convergence in curved space tests based upon
Tolman--Oppenheimer--Volkoff equilibrium solutions for spherical fluids. In the
dynamic self-gravitating case, simulations of spherical accretion of a fluid
onto a black hole were stable and free of numerical problems. Accretion was
successfully carried out in the regime where the mass of the black hole
doubled. Subsequently the code was used to study how accretion modulates both
the decay rates and oscillation frequencies of the quasi-normal modes of the
interior black hole~\cite{imprints}.

The characteristic hydrodynamic approach of Font and Papadopoulos was first applied
to spherical symmetric problems of astrophysical interest. Linke, Font,
Janka, M{\"{u}}ller, and Papadopoulos~\cite{linke} simulated the spherical
collapse of supermassive stars, using an equation of state that included the
effects due to radiation, electron-positron pair formation, and neutrino
emission. They were able to follow the collapse from the onset of instability
to black hole formation. The simulations showed that collapse of a star with
mass greater than $5\times 10^5$ solar masses does not produce enough
radiation to account for the gamma ray bursts observed at cosmological
redshifts.

Next, Siebel, Font, and Papadopoulos~\cite{sieb2002} studied the interaction of
a massless scalar field with a neutron star by means of the coupled
Klein--Gordon--Einstein-hydrodynamic equations. They analyzed the nonlinear scattering
of a compact ingoing scalar pulse incident on a spherical neutron star in an
initial equilibrium state obeying the null version of the
Tolman--Oppenheimer--Volkoff equations. Depending upon the initial mass and
radius of the star, the scalar field either excites radial pulsation modes or
triggers collapse to a black hole. The transfer of scalar energy to the star
was found to increase with the compactness of the star. The approach included
a compactification of null infinity, where the scalar radiation was computed.
The scalar waveform showed quasi-normal oscillations before settling down to a
late time power law decay in good agreement with the $t^{-3}$ dependence
predicted by linear theory. Global energy balance between the star's
relativistic mass and the scalar energy radiated to infinity was confirmed.


\subsection{Axisymmetric characteristic hydrodynamic simulations}
\label{sec:ahydro}

The approach initiated by Font and Papadopoulos has been applied in axisymmetry
to pioneering studies of gravitational waves from relativistic stars. The
gravitational field is treated by the original Bondi formalism using the
axisymmetric code developed by Papadopoulos in his thesis~\cite{papath,papa}.
Because of the twist-free property of the axisymmetry in the original Bondi
formalism, the fluid motion cannot have a rotational component about the axis
of symmetry, i.e.\ the fluid velocity is constrained to the $(r,\theta)$ plane.
In his thesis work, Siebel~\cite{Siebel} extensively tested the combined
hydrodynamic-gravity code in the nonlinear, relativistic regime and demonstrated that
it accurately and stably maintained the equilibrium of a neutron star.

As a first application of the code, Siebel, Font, M{\"{u}}ller, and
Papadopoulos~\cite{sieb2002a} studied axisymmetric pulsations of neutron stars,
which were initiated by perturbing the density and $\theta$-component of
velocity of a spherically symmetric equilibrium configuration. The frequencies
measured for the radial and non-radial oscillation modes of the star were
found to be in good agreement with the results from linearized perturbation
studies. The Bondi news function was computed and its amplitude found to be in
rough agreement with the value given by the Einstein quadrupole formula. Both
computations involve numerical subtleties: The computation of the news
involves large terms which partially cancel to give a small result, and the
quadrupole formula requires computing three time derivatives of the fluid
variables. These sources of computational error, coupled with ambiguity in the
radiation content in the initial data, prevented any definitive conclusions.
The total radiated mass loss was approximately $10^{-9}$ of the total mass.

Next, the code was applied to the simulation of axisymmetric supernova core
collapse~\cite{papadop2003}. A hybrid equation of state was used to mimic
stiffening at collapse to nuclear densities and shock heating during the
bounce. The initial equilibrium state of the core was modeled by a polytrope
with index $\Gamma=4/3$. Collapse was initiated by reducing the  polytropic
index to $1.3$. In order to break spherical symmetry, small perturbations were
introduced into the $\theta$-component of the fluid velocity. During the
collapse phase, the central density increased by 5 orders of magnitude. At
this stage the inner core bounced at supra-nuclear densities, producing an
expanding shock wave which heated the outer layers. The collapse phase was
well approximated by spherical symmetry but non-spherical oscillations were
generated by the bounce. The resulting gravitational waves at null infinity
were computed by the compactified code. After the bounce, the Bondi news
function went through an oscillatory build up and then decayed in an $\ell=2$
quadrupole mode. However, a comparison with the results predicted by the
Einstein quadrupole formula no longer gave the decent agreement found in the
case of neutron star pulsations. This discrepancy was speculated to be due to
the relativistic velocities of $\approx 0.2 c$ reached in the core collapse as
opposed to $10^{-4}c$ for the pulsations. However, gauge effects and numerical
errors also make important contributions which cloud any definitive
interpretation. This is the first study of gravitational wave production by
the gravitational collapse of a relativistic star carried out with a
characteristic code. It is clearly a remarkable piece of work which offers up
a whole new approach to the study of gravitational waves from astrophysical
sources.


\subsection{Three-dimensional characteristic hydrodynamic simulations}
\label{sec:3dhydro}

The PITT code has been coupled with a rudimentary matter source to carry out
three-dimensional characteristic simulations of a relativistic star orbiting a
black hole. Animations can be viewed at~\cite{lsu}. A naive numerical
treatment of the Einstein-hydrodynamic system for a perfect fluid was incorporated
into the code~\cite{matter}, but a more accurate HRSC hydrodynamic algorithm has not
yet been implemented. The fully nonlinear matter-gravity null code was tested
for stability and accuracy to verify that nothing breaks down as long as the
fluid remains well behaved, e.g., hydrodynamic shocks do not form. The code
was used to simulate a localized blob of matter falling into a black hole,
verifying that the motion of the center of the blob approximates a geodesic and
determining the waveform of the emitted gravitational radiation at ${\cal
I}^+$. This simulation was a prototype of a neutron star orbiting a black
hole, although it would be unrealistic to expect that this naive fluid code
would reliably evolve a compact star for several orbits as it spiraled into a
black hole. A 3D HRSC characteristic hydrodynamic code would open the way to
explore this important astrophysical problem.

Short term issues were explored with the code in subsequent
work~\cite{matter2}. The code was applied to the problem of determining
realistic initial data for a star in circular orbit about a black hole. In
either a Cauchy or characteristic approach to this initial data problem, a
serious source of physical ambiguity is the presence of spurious gravitational
radiation in the gravitational data. Because the characteristic approach is
based upon a retarded time foliation, the resulting spurious outgoing waves
can be computed by carrying out a short time evolution. Two very different
methods were used to prescribe initial gravitational null data:
\begin{enumerate}
\item a Newtonian correspondence method, which guarantees that the
  Einstein quadrupole formula is satisfied in the Newtonian
  limit~\cite{newt}, and
  \label{method_1}
\item setting the shear of the initial null hypersurface to zero.
  \label{method_2}
\end{enumerate}
Both methods are
mathematically consistent but suffer from physical shortcomings. Method~\ref{method_1}
has only approximate validity in the relativistic regime of a star in close
orbit about a black hole while Method~\ref{method_2} completely ignores the
gravitational lensing effect of the star. It was found that, independently of
the choice of initial gravitational data, the spurious waves quickly radiate
away, and that the system relaxes to a quasi-equilibrium state with an
approximate helical symmetry corresponding to the circular orbit of the star.
The results provide justification of recent approaches for initializing the
Cauchy problem which are based on imposing an initial helical symmetry, as
well as providing a relaxation scheme for obtaining realistic characteristic
data.


\subsubsection{Massive particle orbiting a black hole}
\label{sec:part}

One attractive way to avoid the computational expense of hydrodynamics in
treating a star orbiting a massive black hole is to treat the star as a
particle. This has been attempted using the PITT code to model a star of mass
$m$ orbiting a black hole of much larger mass, say $1000\,m$~\cite{particle}.
The particle was described by the perfect fluid energy-momentum tensor of a
rigid Newtonian polytrope in spherical equilibrium of a fixed size in its
local proper rest frame, with its center following a geodesic. The validity of
the model requires that the radius of the polytrope be large enough so that
the assumption of Newtonian  equilibrium is valid but small enough so that the
assumption of rigidity is consistent with the tidal forces produced by the
black hole. Characteristic initial gravitational data for a double null
initial value problem were taken to be Schwarzschild data for the black hole.
The system was then evolved using a fully nonlinear characteristic code. The
evolution equations for the particle were arranged to  take  computational
advantage of the energy and angular momentum conservation laws which would
hold in the test body approximation.

The evolution was robust and could track the particle for two orbits as it
spiraled into the black hole. Unfortunately, the computed rate of inspiral was
much too large to be physically realistic: the energy loss was $\approx 10^3$
greater than the value expected from perturbation theory. This discrepancy
might have a physical origin, due to the choice of initial  gravitational data
that ignores the particle or  due to a breakdown of the rigidity assumption,
or a numerical origin due to improper resolution of the particle. It is a
problem whose resolution would require the characteristic AMR techniques being
developed~\cite{pretlehn}.


\subsubsection{Computing the radiation field}

The Bondi news function, which represents the gravitational radiation field at
$\mathcal{I}^+$, is computed by post-processing the output data at $\mathcal{I}^+$
for the primary evolution variables. This is a delicate numerical procedure
involving large terms which partially cancel to give a small result. It is
somewhat analogous to the experimental task of isolating a transverse
radiation field from the longitudinal fields representing the total mass,
while in a very non-inertial laboratory.

In the original algorithm~\cite{high}, the procedure was carried out in the
coordinate system of the code by giving a geometric procedure for computing
the news. This approach has been tested to be second order convergent in a
wide number of testbeds~\cite{high,Zlochower,zlochmode}. Alternatively, a
coordinate transformation may be carried out to inertial Bondi coordinates (as
originally formulated by Bondi~\cite{bondi}), in which the news calculation is
quite clean. This approach was recently implemented in~\cite{bishnews} and
shown to be second order convergent in Robinson--Trautman and Schwarzschild
testbeds. A direct comparison of the two approaches was not made, although it
is clear both face the same delicate numerical problem of extracting a small
radiation field in the presence of large gauge effects in the primary output
data.

The procedure is further complicated by sources of numerical error,
such as
\begin{itemize}
\item the breakdown of the Bondi surface area coordinate $r$ near a
  stationary event horizon,
\item strong nonlinearities and the gauge effects they produce near
  $\mathcal{I}^+$, and
\item sharp gradients.
\end{itemize}

These sources of error can be further aggravated by the introduction of matter
fields, as encountered in trying to make definitive comparisons between the
Bondi news and the Einstein quadrupole formula in the axisymmetric studies of
supernova collapse~\cite{papadop2003} described in Section~\ref{sec:ahydro}. In
the three-dimensional characteristic simulations of a star orbiting a black
hole~\cite{matter2,particle}, the lack of resolution introduced by a localized
star makes an accurate calculation of the news highly problematical. There
exists no good testbed for validating the news calculation in the presence of
a fluid source. A perturbation analysis in Bondi coordinates of the
oscillations of an infinitesimal fluid shell in a Schwarzschild
background~\cite{bishlin} might prove useful for testing constraint
propagation in the presence of a fluid. However, the underlying Fourier mode
decomposition requires the gravitational field to be periodic so that the
solution cannot be used to test computation of mass loss or radiation reaction
effects.

\newpage


\section{Acknowledgments}
\label{acknow}

This work was partially supported by National Science Foundation grant
PHY-0244673 to the University of Pittsburgh. I want to thank the many
people who have supplied me with material. Please keep me updated.

\newpage


\bibliography{refs}

\end{document}